\documentclass[10pt, conference, letterpaper]{IEEEtran}

\IEEEoverridecommandlockouts
\AtBeginDocument{%
  \providecommand\BibTeX{{%
    \normalfont B\kern-0.5em{\scshape i\kern-0.25em b}\kern-0.8em\TeX}}}
    
\usepackage[T1]{fontenc}% optional T1 font encoding
\usepackage{graphicx}
\usepackage{float}
\graphicspath{ {figs/} }
\usepackage{dsfont}
\usepackage{amsmath}
 
\usepackage[ruled,vlined]{algorithm2e}
\usepackage[font=footnotesize]{caption}
\usepackage[font=scriptsize]{subcaption}
\usepackage[export]{adjustbox}
\usepackage{siunitx}
\usepackage{colortbl}
\usepackage{algorithmic}
\usepackage[thinlines]{easytable}
\usepackage{circledsteps}
\usepackage{siunitx}
\usepackage[normalem]{ulem}
\usepackage{amsfonts}
\usepackage{multirow,verbatim}
\usepackage[normalem]{ulem}
\usepackage[dvipsnames]{xcolor}
\usepackage{url}
\usepackage{tikz}
\usetikzlibrary{tikzmark}
\usepackage{multirow}
\usepackage{amsthm}
\usepackage{amsmath}
\usepackage[english]{babel}
\usepackage{tabularx}
\usepackage{booktabs}
\usepackage{array}
\usepackage{threeparttable}
\usepackage{xurl}
\usepackage{hyperref}
\usepackage{titlesec}
\usepackage{bbm}
\titlespacing*{\section}
{0pt}{1.2ex plus 0.2ex minus 0.2ex}{0.6ex}

\titlespacing*{\subsection}
{0pt}{1.0ex plus 0.2ex minus 0.2ex}{0.4ex}

\titlespacing*{\subsubsection}
{0pt}{0.8ex plus 0.2ex minus 0.2ex}{0.3ex}
\newcommand{\probP}{\text{I\kern-0.15em P}}

\usepackage{booktabs,makecell,multirow,hhline,threeparttable} 
\usepackage{xcolor,colortbl,pgf}
\definecolor{Color1}{RGB}{240, 240, 240}
\usepackage{makecell}
\usepackage[top=0.7in, bottom=1.0in, left=0.7in, right=0.7in]{geometry}

\definecolor{lightgray}{RGB}{211,211,211}
\definecolor{LightCyan}{RGB}{224,255,255}

\usepackage{bbm}

\SetKwComment{Comment}{$\triangleright$\ }{}

\newcolumntype{b}{>{\columncolor{blue!10}}c}
\newcolumntype{y}{>{\columncolor{yellow!10}}c}
\newcolumntype{d}{>{\columncolor{red!7}}c}

\SetKwInput{KwInput}{Input}                % Set the Input
\SetKwInput{KwOutput}{Output}              % set the Output

\begin{document}
\title{IRIS: \underline{I}ntelligent \underline{R}andom Access with \underline{I}ntegrated \underline{S}hared Learning for Decentralized LEO Direct-to-Satellite Uplinks} 

%\title{Collision-Aware Decentralized Uplink Scheduling for LEO Direct-to-Satellite Systems}
% \title{CAERUS: \underline{C}ollision-\underline{A}ware Efficient Reinforcement-learning-based Uplink Scheduling for LEO Direct-to-Satellite Systems}

\author{
\IEEEauthorblockN{
\small
Abdullah Al Noman\IEEEauthorrefmark{1},
Gaurav Singh\IEEEauthorrefmark{1},
Rameez Ahmed\IEEEauthorrefmark{2}, and
Debashri Roy\IEEEauthorrefmark{1}
}

\IEEEauthorblockA{
\small
\IEEEauthorrefmark{1}The University of Texas at Arlington, Arlington, TX, USA\\
\IEEEauthorrefmark{2}MathWorks Inc., USA\\
\footnotesize
Emails:
\texttt{abdullahal.noman@uta.edu},
\texttt{gaurav.singh@uta.edu},\\
\texttt{rrasheed@mathworks.com},
\texttt{debashri.roy@uta.edu}
\vspace*{-15pt}
}
}
\maketitle

\begin{abstract}
Grant-free uplink access in low-Earth-orbit (LEO) non-terrestrial networks (NTNs) poses a coordination challenge when multiple UEs share a satellite channel without real-time centralized scheduling under long propagation delays and time-varying channels. Conventional random access does not exploit peer context, leading to collisions and inefficient channel use as contention increases. To address this problem, we propose IRIS, a multi-agent reinforcement learning framework for decentralized grant-free direct-to-satellite (D2S) uplink access, with A2A-IRIS as its sidelink-assisted Agent-to-Agent (A2A) realization. Each UE executes a shared binary transmission policy using local traffic/channel information augmented by compact peer states exchanged over UE-to-UE sidelink, while Vanilla-IRIS serves as a local-only ablation.
Evaluations in rural and suburban LEO D2S deployments use TLE-based trajectories propagated using the Simplified General Perturbations 4
(SGP4), with a Walker-Delta constellation as fallback. A2A-IRIS achieves $2.1579$ and $5.4478$ Mbps, corresponding to $99.9\%$ and $96.1\%$ of the respective scenario-specific upper bounds. Relative to Vanilla-IRIS, throughput improves by $40.8\%$ and $325.0\%$, with collision rates of $0.0000$ and $0.0433$, while one- and two-epoch delayed centralized schedulers are outperformed by approximately $51$--$130\%$. A2A-IRIS also improves fairness, service coverage, and transmission efficiency with moderate Deep Q-Network (DQN) inference overhead. Active-UE load scalability results further show sustained performance
over the evaluated load range. These results demonstrate the effectiveness of sidelink-assisted A2A
coordination for decentralized LEO D2S uplink access under propagation
delay and delayed channel feedback.
\end{abstract}


%%%%%%sectionsasdfsd
% \vspace*{-5pt}
% \input{Sections/Old-Sections/1-old-intro}
\section{Introduction}
\label{sec:intro}
{\color{black}
The growing integration of low-Earth-orbit (LEO) satellite systems with terrestrial cellular networks is expanding the role of non-terrestrial networks (NTNs) in beyond-5G and sixth-generation (6G) communications~\cite{9579443}. In particular, direct-to-device and direct-to-satellite connectivity enables ordinary user equipment (UE) devices to communicate with satellite infrastructure without requiring a terrestrial base station in the service-link path, thereby extending coverage to areas where terrestrial radio access is unavailable, sparse, or temporarily disrupted \cite{10.1145/3697253.3697265}. Among the different NTN architectures, LEO systems are especially advantageous because their lower orbital altitude reduces propagation delay relative to higher-orbit satellite systems while supporting wide-area coverage.

\begin{figure}[t]
\centering
\includegraphics[trim={0.7cm 0.6cm 0.7cm 0.7cm},clip,width=\columnwidth]{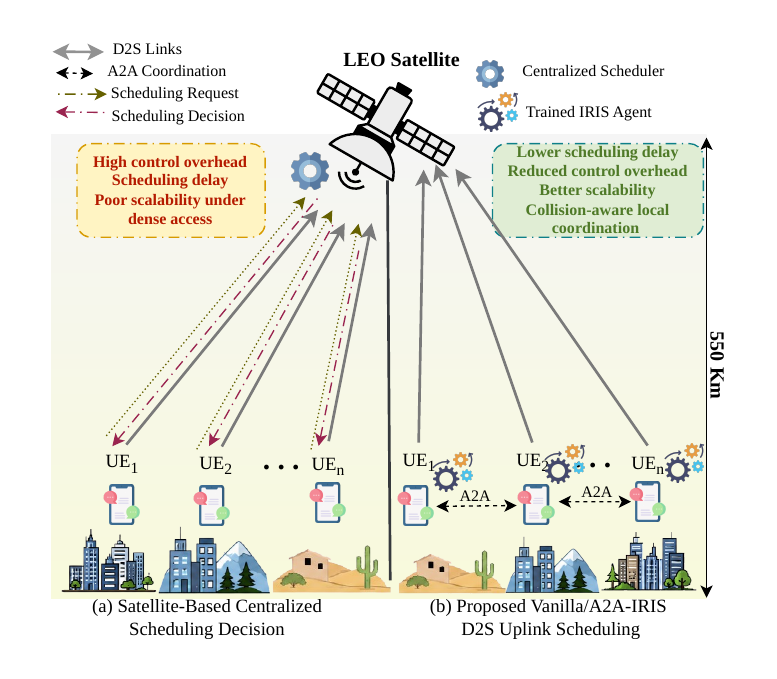}
\caption{Conceptual comparison between centralized LEO D2S uplink scheduling and the proposed IRIS decentralized Agentic AI framework. All UEs act as autonomous agents executing a shared policy using locally available information. The sidelink-assisted configuration additionally exploits Agent-to-Agent (A2A) peer context, whereas the `Vanilla' (local-only) configuration operates without A2A peer-state exchange over the sidelink.}
\vspace{-20pt}
\label{fig:intro-fig}
\end{figure}

\noindent\textbf{Challenges in LEO D2S Uplink Access.} Despite these advantages, efficient uplink access remains challenging. As illustrated in Fig.~\ref{fig:intro-fig}, a large number of geographically distributed UEs may contend for a shared satellite uplink, and poorly coordinated simultaneous transmissions can result in collisions, wasted transmission energy, increased waiting time, and persistent queue buildup. These effects become more pronounced under bursty traffic and finite UE buffers,  where repeated  unsuccessful access attempts directly reduce the useful payload delivered over the satellite link. The access problem is further complicated by the dynamics of the LEO
channel. Rapid satellite motion causes time-varying visibility,
elevation-dependent path loss, propagation delay, and substantial
Doppler shifts, all of which must be handled by the air interface
\cite{ 9579443, 9970355,3gpp_tr38821}. For representative
LEO deployments, 3GPP studies report service-link delays on the order of
several milliseconds together with Doppler shifts reaching tens of
kilohertz at S-band frequencies~\cite{3gpp_tr38821}. Although centralized scheduling remains possible in NTN systems, these
propagation and channel dynamics increase the staleness of channel-state
information (CSI) and the control-loop overhead required for rapid
per-epoch decisions based on fresh network-wide channel information. The increasing complexity and scale of these dynamic telecommunications environments demand intelligent automation, where Agentic AI has emerged as a pivotal paradigm enabling autonomous agents to perceive, reason, act, and continuously learn to optimize network configurations. This motivates access mechanisms that transition decision-making directly to the UE edge, relying on distributed intelligence rather than strictly centralized control loops.

\noindent\textbf{Limitations of Existing NTN Access Mechanisms.} 3GPP has introduced several NR-NTN mechanisms for satellite operation.
Release~17 includes NTN-specific timing management, GNSS-assisted Doppler
pre-compensation, mobility support, and HARQ adaptations
\cite{9579443,liu2022uplink, 3gpp2022ts38300}. These mechanisms
address synchronization, mobility, and PHY/MAC operation over long and
time-varying satellite links, but do not directly determine how multiple
continuously backlogged UEs should autonomously coordinate repeated access
to a shared uplink contention domain.

Prior work has addressed complementary NTN random-access issues,
including two-stage detection for LEO NTN IoT access
~\cite{10158920} and reinforcement learning (RL) and multi-agent
reinforcement learning (MARL)-based NTN resource allocation~\cite{10018208}.
More broadly, Agentic AI extends multi-agent learning toward
self-organizing and adaptive network operation. However, decentralized
Agentic AI for handheld UEs making binary access decisions over a single
shared D2S contention domain under partial observability remains
insufficiently explored.

\noindent\textbf{Proposed IRIS Framework.}
To address this problem, we propose \emph{IRIS}, an Agentic AI-driven
random-access framework based on parameter-shared MARL for decentralized
LEO D2S uplinks. As illustrated in Fig.~\ref{fig:intro-fig},
conventional satellite-based scheduling relies on request/decision
exchanges over the D2S path, incurring control overhead and delay. IRIS
instead moves access decisions to the UE side, where each trained agent
independently selects its uplink action. A2A-IRIS further exploits
sidelink-based Agent-to-Agent (A2A) coordination, allowing compact peer
information to assist contention resolution without per-epoch satellite scheduling.

All UEs share a DQN with binary action space
$\{\mathrm{NT},\mathrm{T}\}$, where $\mathrm{NT}$ and $\mathrm{T}$
denote no transmission and transmission, respectively. Parameter sharing
keeps the policy parameterization independent of the UE population.
A2A-IRIS is the proposed sidelink-assisted realization, while
Vanilla-IRIS is its local-only ablation. Vanilla-IRIS observes normalized
queue occupancy, effective SINR, waiting time, previous access outcome,
and a UE-specific coordination token. A2A-IRIS additionally uses a
fixed-dimensional histogram of delayed peer queue, channel, and waiting
states, together with current relative contention and  predicted-payload ranks.

Peer exchange leverages the existing 3GPP PC5 sidelink
interface~\cite{9461188}, rather than introducing a new
coordination interface. Although conceptually related to Google's
general-purpose A2A agent-interoperability paradigm
~\cite{googleA2A2025}, IRIS uses a lightweight, domain-specific A2A
mechanism for D2S random access. Peer information does not
deterministically select a transmitter; each UE independently evaluates
the shared DQN and chooses $\mathrm{NT}$ or $\mathrm{T}$. Thus, the
Vanilla-IRIS comparison isolates the benefit of peer context under the
same action space, reward, PHY, traffic, and learning framework.

During training, load-aware exploration limits excessive random
transmissions, event-preserving replay retains informative transmission
events, and action-aware minibatch sampling reduces domination by
$\mathrm{NT}$ transitions. A cooperative reward combines UE-specific
service utility with a common access-outcome term to discourage
collisions, idle epochs with backlog, and singleton PHY failures while
promoting urgent service. During inference, the DQN is frozen and each
UE executes greedily without replay insertion, gradient updates, or
target-network synchronization.

\noindent\textbf{Our Contributions.} The main contributions are: 
% summarized as follows:

\noindent \textbf{(1)} \textbf{Parameter-Shared Decentralized Agentic AI for D2S Random Access:}
We develop IRIS, an Agentic AI framework in which homogeneous UEs act as autonomous agents that share a common DQN and independently select between  $\mathrm{NT}$ and $\mathrm{T}$ over a single shared D2S uplink contention domain. Parameter sharing avoids
per-UE networks and enables decentralized online execution from local
observations.

\noindent \textbf{(2)} \textbf{Agent-to-Agent (A2A) Sidelink-Assisted Observation Configurations:}
We propose A2A-IRIS as the sidelink-assisted realization and Vanilla-IRIS
as its local-only ablation. Both use local queue, effective-SINR,
waiting, previous-outcome, and contention-token information, while
A2A-IRIS additionally incorporates fixed-dimensional peer context.
Delayed peer states are compressed into a normalized histogram and
combined with a current contention-priority rank, keeping the DQN input
dimension independent of the UE population.

\noindent \textbf{(3)} \textbf{Access-Aware Deep-RL Training and Cooperative Reward:}
We design training for the imbalanced binary-access problem using
load-aware exploration, event-preserving replay, and action-balanced
mini-batches to control random attempts, retain informative TX
experiences, and limit domination by silence transitions. The reward
combines delivered payload, queue/waiting urgency, and a cooperative
access-outcome term to discourage collisions, unnecessary idling, PHY
failures, and persistent UE capture.

\noindent \textbf{(4)} \textbf{Population- and Geometry-Driven LEO D2S Evaluation:}
We evaluate IRIS using population-density-driven UE populations,
pedestrian traffic and mobility, and TLE-based trajectories
~\cite{celestrak} propagated with SGP4~\cite{vallado2006sgp4}.
The evaluation further incorporates a 3GPP-inspired LEO S-band channel
with elevation-dependent propagation, Doppler compensation, MCS
adaptation, BLER, and HARQ under suburban and rural conditions. A2A-IRIS is compared with Vanilla-IRIS, a centralized upper
bound, current and one-/two-epoch delayed channel-quality-based centralized schedulers,
p$^\star$-ALOHA, and binary exponential backoff. Performance is evaluated
using throughput, access outcomes, fairness, RF energy and transmission
resource costs, DQN inference complexity, and active-UE load scalability.

% \end{enumerate}
% }
% The remainder of this paper is organized as follows. Section~II reviews related work and motivation. Section~III presents the D2S system model and problem formulation. Section~IV introduces the proposed IRIS framework, including the Vanilla-IRIS and A2A-IRIS configurations. Section~V describes the experimental setup, baselines, and evaluation methodology. Section~VI presents the performance evaluation, and Section~VII concludes the paper.

% \input{Sections/1-introduction}

\vspace{4pt}
\section{Related Work and Motivation}
\label{sec:related_works}
\noindent {\bf NTN and LEO Satellite Architectures.}
Large LEO constellations have accelerated the integration of satellite
access with terrestrial cellular systems and strengthened the role of
NTNs in beyond-5G/6G networks~\cite{9210567}. Lin \textit{et al.}~\cite{9579443}
discussed 3GPP NTN standardization challenges in synchronization,
mobility, timing advance, and random access. Recent studies have also
explored multi-layer D2C architectures, including spectrum-intelligent
multi-tier LEO association and UAV-assisted space-air-ground relaying
for improving resource utilization and link reliability
~\cite{noman2026orbit,11310652}. These studies highlight that LEO access
must accommodate time-varying geometry, mobility, and control-plane
dynamics that differ substantially from static terrestrial RANs.

\noindent {\bf Random Access and Synchronization in NTN.}
LEO random access is affected by differential propagation delay,
Doppler, and geographically distributed terminals sharing satellite
coverage. Chougrani \textit{et al.}~\cite{9590534} studied NB-IoT NTN
preamble detection and uplink synchronization under these conditions.

Recent work has also addressed grant-free massive access. Shen
\textit{et al.}~\cite{9849120} proposed massive-MIMO OTFS access with
joint activity detection and channel estimation; Ying \textit{et al.}
~\cite{10.1109/JSAC.2023.3273699} considered cooperative
multi-satellite quasi-synchronous access; Shen \textit{et al.}
~\cite{10.1109/JSAC.2024.3460064} extended cooperative OTFS access to
joint device identification, channel estimation, and symbol detection;
and Zhang \textit{et al.}~\cite{10680640} investigated grant-free NOMA
with joint user identification, channel estimation, and data detection.
These approaches primarily improve waveform- and receiver-level access
reliability, whereas our focus is decentralized UE-side control of
whether to contend for the shared D2S uplink.

\noindent {\bf Collision Detection and Access Robustness.}
Jeong \textit{et al.}~\cite{10938640} proposed early preamble-collision
detection for LEO NTN IoT access to accelerate collision recovery, while
two-stage preamble detection has also been studied for robust LEO NTN
IoT access~\cite{10158920}. Such methods improve collision detection and
recovery after access begins; in contrast, our objective is to reduce
simultaneous transmissions through decentralized UE-side decisions.

\noindent {\bf Learning-Based Random Access and Resource Control.}
Learning-based access control has been explored for adaptive random
access. Jiang \textit{et al.}~\cite{9174918} used DRL to tune access
parameters under varying IoT traffic, while Lee \textit{et al.}
~\cite{9844988} proposed eRACH for decentralized LEO contention without
inter-agent communication. Yu \textit{et al.}~\cite{10132304} considered
DRL-based NOMA slotted ALOHA with repetition and SIC, whereas Oh
\textit{et al.}~\cite{10.1145/3704413.3764464} studied decentralized
MARL with limited peer exchange in terrestrial settings. Queue-aware distributed access, exemplified by Q-CSMA, and CTDE-based MARL, exemplified by QMIX, provide complementary approaches to decentralized coordination~\cite{5462229,3455716.3455894}. In contrast, A2A-IRIS uses a shared
binary access policy with fixed-dimensional peer-assisted observations
for LEO D2S uplink contention.

\noindent {\bf Research Gap and Motivation.}
Despite prior work on NTN architectures, synchronization, random-access
robustness, collision handling, and learning-based control, decentralized
UE-side contention for sustained LEO D2S uplink traffic remains
underexplored. Existing studies largely emphasize PHY-layer access, receiver-side
processing, centralized coordination, richer resource allocation, or
terrestrial decentralized access settings whose signaling and
carrier-sensing assumptions do not directly match LEO D2S operation. IRIS addresses this gap through parameter-shared MARL with a
common DQN for binary $\{\mathrm{NT},\mathrm{T}\}$ access, evaluated
under local-only and sidelink-assisted observations to isolate the value
of compact peer-assisted coordination.
% \input{Sections/Old-Sections/2-old-related}
% \input{Sections/2-related_works}
% \input{Sections/Old-Sections/3-old-problem}

%%%%%%%%%%%%%%%%%%%%%%%%%%%%%%%%%%%%%%%%%%%%%%%%%%%%%%%%%%%%%%%%%%%%%%%%%%%%%%%%
\providecommand{\TX}{\mathrm{T}}
\providecommand{\NT}{\mathrm{NT}}

% =========================================================
% SYSTEM MODEL AND PROBLEM FORMULATION
% =========================================================

\section{System Model and Problem Formulation}
\label{sec:system_model}

\subsection{LEO D2S Network and Population-Driven Traffic Model}
\label{subsec:network_traffic}

We consider a LEO D2S uplink with a set
$\mathcal{U}=\{1,\ldots,U\}$ indexing the active UEs, which communicate
directly with a serving LEO satellite over a dedicated FDD uplink
carrier. Satellite positions and velocities are obtained from
two-line elements (TLEs) propagated using SGP4, with a Walker-Delta
constellation retained as a fallback when TLE-based geometry is
unavailable.

At decision epoch $t$, UE~$i$ considers satellites satisfying
\begin{equation}
\vspace*{-3pt}
\theta_{i,s}(t)\geq\theta_{\min},
\end{equation}
where $\theta_{i,s}(t)$ is the elevation angle of satellite $s$ as
observed by UE~$i$, and $\theta_{\min}$ is the minimum visibility
elevation. UE~$i$ then associates with the candidate providing the
largest desired-link reference received power. Thus, satellite
association evolves with orbital geometry and the large-scale link
budget. The satellite receive beam is directed toward the modeled
service region.

The active UE population is coupled to the deployment environment.
For a square service region of side length $L_{\mathrm{svc}}$ and
population density $\rho_{\mathrm{pop}}$,
\begin{equation}
\vspace*{-3pt}
U
=
\max
\left\{
1,\,
\operatorname{round}
\left(
\rho_{\mathrm{pop}}
L_{\mathrm{svc}}^2
\eta_{\mathrm{cand}}
\eta_{\mathrm{act}}
\right)
\right\},
\label{eq:population_ue_count}
\end{equation}
where $\eta_{\mathrm{cand}}$ is the population-to-candidate-handheld
mapping and $\eta_{\mathrm{act}}$ is the activity factor. Hence, the
contention load follows from the deployment density and service area
rather than being independently selected. Scenario-specific values are
given in Section~\ref{subsec:experiment_setting}.

UEs follow pedestrian mobility with independently drawn headings, which
contributes to the relative satellite--UE velocity and instantaneous
Doppler shift. Each active UE is offered traffic at bit-arrival rate $\lambda$, with the number of arriving bits per epoch modeled as
\begin{equation}
\vspace*{-4pt}
A_i(t)
\sim
\operatorname{Poisson}
\left(
\lambda\Delta t
\right),
\label{eq:poisson_arrivals}
\end{equation}
where $\Delta t$ is the MARL control epoch rather than an NR physical slot.

Each UE maintains a finite queue of capacity $B_{\max}$. At the beginning
of epoch $t$, newly generated uplink traffic $A_i(t)$ is added to UE
$i$'s existing backlog $B_i(t)$ before channel observation and access
selection. Hence, the post-arrival, pre-service queue is
\vspace{-6pt}\begin{equation} \vspace*{-4pt}
\widetilde B_i(t)
=
\min
\left\{
B_i(t)+A_i(t),
B_{\max}
\right\}.
\label{eq:post_arrival_buffer}
\end{equation}
If $D_i(t)$ payload bits are successfully delivered during epoch $t$,
the queue evolves as
\vspace{-6pt}\begin{equation} \vspace*{-4pt}
B_i(t+1)
=
\max
\left\{
\widetilde B_i(t)-D_i(t),
0
\right\}.
\label{eq:buffer_dynamics}
\end{equation}

Thus, queue evolution jointly reflects traffic arrivals, decentralized
access decisions, contention outcomes, and the time-varying D2S link;
silence, collisions, and PHY failures increase queue pressure, whereas
successful singleton transmission reduces the accumulated backlog.

% =========================================================
% ACCESS MODEL
% =========================================================

\subsection{Contention-Based Grant-Free Uplink Access Model}
\label{subsec:access_model}

All backlogged UEs contend over a single shared grant-free uplink
domain. At each decision epoch $t$, UE~$i$ selects
\vspace{-6pt}\begin{equation}\vspace*{-4pt}
a_i(t)\in\mathcal{A}_i=\{\mathrm{NT},\mathrm{T}\},
\label{eq:binary_action}
\end{equation}
where $\mathrm{NT}$ and $\mathrm{T}$ denote no transmission and
transmission, respectively; empty-buffer UEs always select $\mathrm{NT}$.
Let
\begin{equation}\vspace*{-4pt}
N_{\mathrm T}(t)
=
\sum_{i\in\mathcal U}
\mathbbm{1}\!\left\{
a_i(t)=\mathrm T,\,
\widetilde B_i(t)>0
\right\}
\label{eq:number_tx}
\end{equation}
denote the number of transmitting UEs.

If $N_{\mathrm T}(t)=0$, the epoch is idle; if
$N_{\mathrm T}(t)=1$, the singleton UE is evaluated by the D2S PHY and
uses the complete modeled UL allocation of 24 PRBs; and if
$N_{\mathrm T}(t)>1$, a MAC-layer collision occurs with no recovered
payload. Hence, contention failures are distinguished from singleton
PHY failures caused by visibility, insufficient effective SINR, or
stochastic decoding. The contention domain is modeled globally; hence, simultaneous transmissions are treated as a collision even if the involved UEs
associate with different serving satellites.

Let $\chi_i(t)\in\{0,1\}$ denote the singleton decoding outcome and
$C_i(t)$ its PHY service opportunity. The delivered payload is
\begin{equation}\vspace*{-4pt}
\begin{aligned}
D_i(t)
&=
\mathbbm{1}\!\left\{
a_i(t)=\mathrm T,\,
N_{\mathrm T}(t)=1,\,
\chi_i(t)=1
\right\}
\\
&\quad\times
\min\!\left\{
\widetilde B_i(t),C_i(t)
\right\}.
\end{aligned}
\label{eq:delivered_bits}
\end{equation}

The common previous access outcome is
\begin{equation}\vspace*{-4pt}
o(t)=
\begin{cases}
0, & \text{idle or unsuccessful singleton},\\
1, & \text{successful singleton},\\
2, & \text{collision}.
\end{cases}
\label{eq:outcome}
\end{equation}

The policies do not enforce a deterministic single-winner rule; they
learn to favor useful singleton access while reducing collisions and
unnecessary idling. This single-contention-domain model intentionally
isolates decentralized access coordination and does not represent a
complete NR scheduler with parallel PRB allocation, frequency-domain
resource selection, or multi-user decoding.
% =========================================================
% PHY
% =========================================================

\subsection{Channel and PHY Model}
\label{subsec:channel_model}

The D2S PHY accounts for orbital geometry, free-space propagation,
environment-dependent attenuation, satellite receive-beam gain,
shadowing, small-scale fading, Doppler, residual carrier-frequency
offset, inter-carrier interference (ICI), and thermal noise.
For UE~$i$, the large-scale propagation loss is
\begin{equation} \vspace*{-4pt}
L_i(t)
=
L_{\mathrm{FS},i}(t)
+
L_{\mathrm{tropo},i}(t)
+
L_{\mathrm{iono},i}(t)
+
A_{\mathrm{rain},i}(t),
\label{eq:large_scale_loss}
\end{equation}
and the received power is
\begin{equation} \vspace*{-4pt}
\begin{aligned}
P_{r,i}(t)[\mathrm{dBW}]
={}&
\mathrm{EIRP}_i
-
L_i(t)
-
L_{\mathrm{cl},i}(t)
\\
&+
G_r\!\left(\phi_i(t)\right)
+
G_{\mathrm{fad},i}(t)
+
G_{\mathrm{sh},i}(t),
\end{aligned}
\label{eq:received_power}
\end{equation}
where $L_{\mathrm{cl},i}(t)$ denotes NLOS clutter loss,
$G_r(\phi_i(t))$ the satellite receive-beam gain,
$G_{\mathrm{fad},i}(t)$ the small-scale fading gain, and
$G_{\mathrm{sh},i}(t)$ the shadow-fading term.

The receive beam follows a circular-aperture model and is directed
toward the service-area center. Suburban and rural LOS probability,
LOS/NLOS shadowing, and NLOS clutter loss are selected from the
elevation-dependent S-band parameters in 3GPP TR~38.811
\cite{3gpp_tr38811}, using the nearest available reference elevation.
LOS links follow an elevation-dependent Rician model, while NLOS links
use Rayleigh fading, with an NTN-TDL-like multipath profile. Temporal
fading correlation is modeled as
\vspace{-8pt}\begin{equation} \vspace*{-4pt}
\rho_{f,i}(t)
=
J_0
\left(
2\pi
|f_{d,i}(t)|
\Delta t
\right),
\label{eq:fading_correlation}
\end{equation}
where $f_{d,i}(t)$ is the geometry-dependent Doppler shift.

Other visible satellites are not modeled as uplink interferers;
simultaneous UE transmissions are handled by the MAC collision rule.
Thus, singleton transmissions are noise limited before the residual
Doppler-induced ICI penalty. We further assume GNSS-aided Doppler
pre-compensation, with the residual frequency error modeled as
\begin{equation} \vspace*{-4pt}
\delta f_i(t)
=
\alpha_c
\delta f_i(t-1)
+
\sqrt{1-\alpha_c^2}\,
\sigma_c z_i(t),
\qquad
z_i(t)\sim\mathcal{N}(0,1),
\label{eq:residual_doppler}
\end{equation}
where $\alpha_c$ and $\sigma_c$ control temporal correlation and
residual uncertainty, respectively. The normalized residual offset is
\begin{equation} \vspace*{-4pt}
\epsilon_i(t)
=
\frac{|\delta f_i(t)|}{\Delta f},
\label{eq:normalized_cfo}
\end{equation}
with useful-signal and ICI fractions
\begin{align}
S_{\mathrm{sig},i}(t)
&=
\operatorname{sinc}^{2}\!\left(\epsilon_i(t)\right),
\label{eq:useful_signal_fraction}
\\
I_{\mathrm{ICI},i}(t)
&=
1-S_{\mathrm{sig},i}(t).
\label{eq:ici_fraction}
\end{align}

Let $\gamma_i(t)$ denote the nominal singleton-link SINR in linear scale. The effective SINR after the residual-ICI penalty is
\begin{equation} \vspace*{-4pt}
\gamma_i^{\mathrm{eff}}(t)
=
\left[
\gamma_i^{-1}(t)
+
\frac{
I_{\mathrm{ICI},i}(t)
}{
S_{\mathrm{sig},i}(t)
}
\right]^{-1}.
\label{eq:effective_sinr}
\end{equation}

The effective SINR is mapped to an MCS-dependent spectral efficiency
$\eta_i^{\mathrm{SE}}(t)$ and BLER. A bounded HARQ abstraction increases
the effective decoding SINR across retransmission rounds until either
the target BLER or maximum HARQ-round limit is reached. If
$N_{r,i}(t)$ denotes the resulting number of rounds, the singleton
service opportunity is
\begin{equation} \vspace*{-4pt}
C_i(t)
=
\frac{
\eta_i^{\mathrm{SE}}(t)
B_{\mathrm{UL}}
\Delta t
\eta_{\mathrm{d}}
}{
N_{r,i}(t)
},
\label{eq:service_opportunity}
\end{equation}
where $B_{\mathrm{UL}}$ is the occupied uplink bandwidth and
$\Delta t=0.5$~s is a MARL control epoch rather than a physical NR slot,
thereby aggregating multiple NR transmission opportunities. The factor
$\eta_{\mathrm d}$ captures propagation-delay efficiency.

HARQ is abstracted through the effective round count $N_{r,i}(t)$,
which reduces the service in \eqref{eq:service_opportunity}. Hence, the
model captures retransmission-induced throughput loss without explicitly
simulating slot-level ACK/NACK timing or HARQ feedback latency. The final
decoding outcome $\chi_i(t)$ is drawn from the residual BLER.

% =========================================================
% POMARL FORMULATION
% =========================================================

\subsection{Cooperative Partially Observable MARL Formulation}
\label{subsec:problem_formulation}

The decentralized access problem is modeled as a cooperative partially
observable stochastic game. At decision epoch $t$, the latent global
state is
\begin{equation} \vspace*{-4pt}
\mathbf{X}(t)
=
\left\{
\widetilde B_i(t),W_i(t),\gamma_i^{\mathrm{eff}}(t)
\right\}_{i=1}^{U}
\cup
\left\{
o(t-1),\mathcal{G}(t)
\right\},
\label{eq:global_state}
\end{equation}
where $W_i(t)$ is the service waiting age and $\mathcal{G}(t)$ captures
the time-varying orbital and propagation geometry.

Each UE observes only its local state $\mathbf{s}_i(t)$ and selects an
action independently, yielding
\begin{equation} \vspace*{-4pt}
\mathbf{a}(t)=[a_1(t),\ldots,a_U(t)].
\end{equation}
The transition
\begin{equation} \vspace*{-4pt}
\mathbf{X}(t+1)
\sim
\mathcal{P}\!\left(
\cdot\mid\mathbf{X}(t),\mathbf{a}(t)
\right)
\end{equation}
is driven by traffic arrivals, queue service, satellite/UE motion,
fading, residual Doppler error, and contention outcomes.

All UEs share a common DQN parameterization. Let
$\pi_{\boldsymbol{\theta}}(a_i|\mathbf{s}_i)$ denote the shared
decentralized policy. The objective is
\begin{equation} \vspace*{-4pt}
\boldsymbol{\theta}^{\star}
=
\arg\max_{\boldsymbol{\theta}}
\mathbb{E}
\left[
\sum_{t=0}^{\infty}
\gamma_{\mathrm{RL}}^{t}
\frac{1}{U}
\sum_{i=1}^{U}
r_i(t)
\right],
\label{eq:final_problem}
\end{equation}
where $\gamma_{\mathrm{RL}}\in(0,1)$ is the RL discount factor; reward
averaging avoids trivial scaling with the UE population.

Each UE has partial information and interacts with others through the
shared collision domain. In A2A-IRIS, compact coordination information
is exchanged over an idealized collision-free UE-to-UE sidelink before
the D2S access decision. Reliable delivery within the
$\Delta t=0.5$~s control epoch is assumed, while sidelink PHY/MAC
resource allocation, packet loss, interference, and retransmissions are
abstracted. Delayed peer states use the preceding epoch, whereas current
contention and payload-priority features are available before the present
$\{\mathrm{NT},\mathrm{T}\}$ decision.

% =========================================================
% PROPOSED FRAMEWORK
% =========================================================

\section{Proposed IRIS Framework}
\label{sec:proposed}

As illustrated in Fig.~\ref{fig:iris_architecture}, we develop a
parameter-shared deep-RL framework for decentralized LEO D2S random
access. A2A-IRIS is the sidelink-assisted realization, while
Vanilla-IRIS serves as its local-only ablation. Both use the same binary
action space $\{\mathrm{NT},\mathrm{T}\}$, reward, collision model, and
training procedure, differing only in the availability of peer-assisted
coordination.

All UEs share the action-value function
\begin{equation}\vspace*{-4pt}
Q_{\boldsymbol{\theta}}:
\mathcal{S}\times\mathcal{A}
\rightarrow\mathbb{R},
\end{equation}
with each UE selecting its action from its own observation.

\begin{figure*}[t!]
    \centering
    \includegraphics[
        width=\textwidth,
        trim=0 0mm 0 0mm,
        clip
    ]{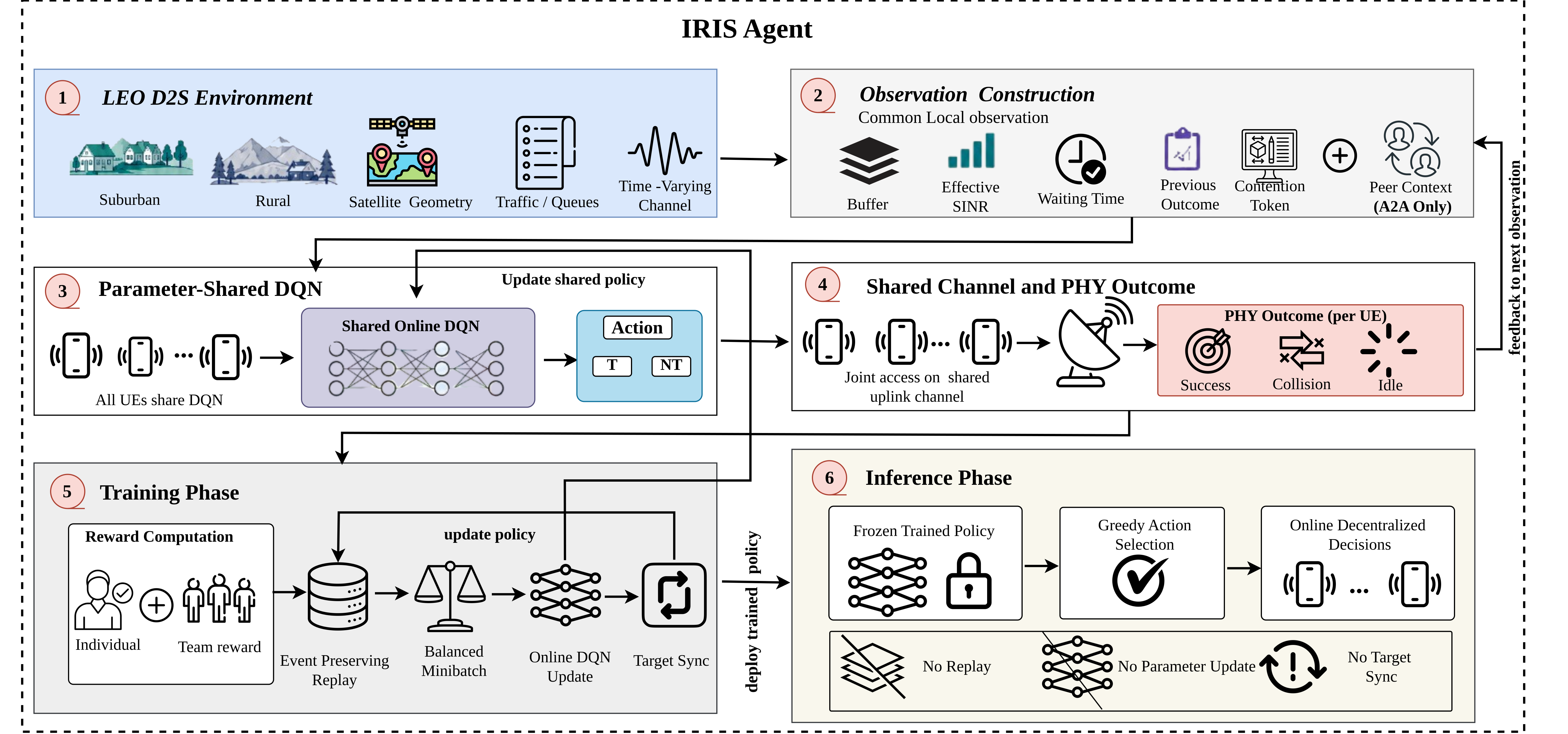}
    \caption{IRIS framework for decentralized LEO D2S uplink scheduling,
    showing observation construction, parameter-shared DQN access decisions,
    shared-channel outcomes, centralized training with replay and target
    updates, and decentralized greedy inference using a frozen policy.}
    \label{fig:iris_architecture}
    \vspace{-15pt}
\end{figure*}
% =========================================================
% OBSERVATION
% =========================================================
\subsection{Local Observation and Sidelink-Assisted Peer Context}
\label{subsec:state_action}

Both configurations share the five-dimensional local observation
\begin{equation} \vspace*{-4pt}
\mathbf{x}_i(t)
=
\left[
\bar B_i(t),\,
\bar\gamma_i(t),\,
\bar W_i(t),\,
\frac{o(t-1)}{2},\,
\xi_i(t)
\right]^{\mathsf T},
\label{eq:local_features}
\end{equation}
where the normalized decision-time queue occupancy and effective-SINR
features are
\begin{equation} \vspace*{-4pt}
\bar B_i(t)
=
\operatorname{clip}
\left(
\frac{\widetilde B_i(t)}{B_{\max}},
0,1
\right),
\label{eq:buffer_feature}
\end{equation}
and
\begin{equation} \vspace*{-4pt}
\bar\gamma_i(t)
=
\operatorname{clip}
\left(
\frac{
\gamma_i^{\mathrm{eff}}(t)-\gamma_{\min}
}{
\gamma_{\max}-\gamma_{\min}
},
0,1
\right),
\label{eq:sinr_feature}
\end{equation}
respectively. The waiting age is mapped smoothly to the unit interval as
\begin{equation} \vspace*{-4pt}
\bar W_i(t)
=
1-
\exp
\left(
-\frac{W_i(t)}{\tau_W}
\right),
\label{eq:wait_feature}
\end{equation}
where $\tau_W$ controls the saturation rate.

The final local feature is a stochastic contention token. UE~$i$ draws
% % \begin{equation}
$u_i(t)\sim\mathcal U(0,1),$
% \end{equation}
and forms
\begin{equation} \vspace*{-4pt}
\xi_i(t)=
\frac{\log(1+Uu_i(t))}
     {\log(1+U_{\max})}.
\label{eq:contention_token}
\end{equation}
where $U_{\max}$ is the maximum supported UE population used for
normalization. The token does not directly determine access; it provides
a private symmetry-breaking variable for UEs with otherwise similar
local observations.

\subsubsection{Compact Delayed Peer-State Representation}

For sidelink-assisted coordination, each UE quantizes its buffer, channel,
and waiting states into three levels,
\begin{equation}\vspace*{-4pt}
b_i(t),\;
c_i^{\mathrm{bin}}(t),\;
w_i^{\mathrm{bin}}(t)
\in\{1,2,3\}.
\end{equation}
These states are packed into a 27-level code,
\begin{equation*}\vspace*{-4pt}
\widetilde p_i(t)
=
9\!\left(b_i(t)-1\right)
+3\!\left(c_i^{\mathrm{bin}}(t)-1\right)
+\left(w_i^{\mathrm{bin}}(t)-1\right)+1,
\label{eq:peer_code}
\end{equation*}

Peer-state information is delayed by one decision epoch, so the
observation at $t$ uses messages generated at $t-1$. Let
$\mathcal{V}_i(t-1)$ denote the peers with valid delayed codes. The codes
are compressed into a fixed-dimensional normalized histogram:
\begin{equation}\vspace*{-4pt}
 \resizebox{\linewidth}{!}{ $ h_{i,k}(t)=
\begin{cases}
\dfrac{\displaystyle\sum_{j\in\mathcal{V}_i(t-1)}
\mathbbm{1}\!\left\{\widetilde p_j(t-1)=k\right\}}
{|\mathcal{V}_i(t-1)|},
&|\mathcal{V}_i(t-1)|>0,\\[6pt]
0,&|\mathcal{V}_i(t-1)|=0,
\end{cases}
\quad k=1,\ldots,27,
$}
\label{eq:peer_histogram}
\end{equation}
with
\begin{equation}\vspace*{-4pt}
\mathbf h_i(t)=
[h_{i,1}(t),\ldots,h_{i,27}(t)]^{\mathsf T}.
\label{eq:peer_histogram_vector}
\end{equation}
Thus, the representation preserves the empirical distribution of peer
states while remaining invariant to UE ordering and independent of $U$.

\subsubsection{Current Contention and Payload Ranks}

In addition to the delayed peer histogram, A2A-IRIS uses two current
relative-priority features. At the beginning of decision epoch $t$,
each backlogged UE shares its contention token $\xi_i(t)$ and predicted
queue-limited payload $\widehat D_i(t)$ through the sidelink before the
TX/NT decision. Here, $\widehat D_i(t)$ denotes the expected queue-limited payload
computed from the current local PHY estimate, residual BLER, and
available backlog. Define
\begin{equation}\vspace*{-4pt}
\mathcal{A}_B(t)
=
\left\{
j:\widetilde B_j(t)>0
\right\}.
\label{eq:active_backlog_set}
\end{equation}
For $i\in\mathcal{A}_B(t)$, the corresponding percentile ranks are
\begin{align}
\rho_i^{\xi}(t)
&=
\frac{
\sum_{j\in\mathcal{A}_B(t)}
\mathbbm{1}\!\left\{\xi_j(t)\leq\xi_i(t)\right\}
}{
|\mathcal{A}_B(t)|
},
\label{eq:contention_rank}
\\
\rho_i^{D}(t)
&=
\frac{
\sum_{j\in\mathcal{A}_B(t)}
\mathbbm{1}\!\left\{
\widehat D_j(t)\leq\widehat D_i(t)
\right\}
}{
|\mathcal{A}_B(t)|
}.
\label{eq:payload_rank}
\end{align}
For an empty-buffer UE, both ranks are set to zero. Unlike the
one-epoch-delayed peer histogram, these ranks use coordination
information available within the current control epoch. They provide
relative priority only and do not impose a deterministic transmitter
selection rule.

The complete sidelink-assisted observation is
\begin{equation}\vspace*{-8pt}
\mathbf{s}^{\mathrm{SL}}_i(t)
=
\left[
\mathbf{x}_i^{\mathsf T}(t),
\mathbf{h}_i^{\mathsf T}(t),
\rho_i^\xi(t),
\rho_i^D(t)
\right]^{\mathsf T}.
\label{eq:sl_state}
\end{equation}
Its dimension therefore remains independent of the UE population.

% =========================================================
% SHARED DQN
% =========================================================

\subsection{Parameter-Shared Deep Q-Learning}
\label{subsec:dqn_training}

For observation $\mathbf{s}_i(t)$, the shared DQN produces the two possible
actions
\begin{equation} \vspace*{-8pt}
\mathbf{q}_i(t)
=
\left[
Q_{\boldsymbol{\theta}}
\left(
\mathbf{s}_i(t),\NT
\right),
\;
Q_{\boldsymbol{\theta}}
\left(
\mathbf{s}_i(t),\TX
\right)
\right].
\label{eq:q_values}
\end{equation}

The network consists of a feed-forward input layer, two fully connected
hidden layers with nonlinear activation, and a two-dimensional output
layer. The layer widths and training hyperparameters are provided in
Section~\ref{subsec:experiment_setting}.

For a replay-buffer transition tuple
$\left(\mathbf{s},a,r,\mathbf{s}',d\right)$,
where $d\in\{0,1\}$ is the terminal-state indicator, the target network
$Q_{\boldsymbol{\theta}^{-}}$ computes the one-step DQN target as
\begin{equation} \vspace*{-8pt}
y=r+\gamma_{\mathrm{RL}}(1-d)
\max_{a'\in\mathcal A}Q_{\theta^-}(s',a').
\label{eq:dqn_target}
\end{equation}

The online network minimizes
\begin{equation} \vspace*{-8pt}
\mathcal{L}
\left(
\boldsymbol{\theta}
\right)
=
\mathbb{E}
\left[
\left(
Q_{\boldsymbol{\theta}}
\left(
\mathbf{s},a
\right)
-y
\right)^2
\right].
\label{eq:dqn_loss}
\end{equation}

A separate target network is periodically synchronized with the online
network to reduce instability caused by simultaneously changing
bootstrapping targets.

% =========================================================
% REPLAY
% =========================================================

\subsubsection{Event-Preserving Replay}

The one-winner contention model naturally produces substantially more
$\NT$ decisions than $\TX$ decisions. Uniformly storing and sampling
all local observations would therefore allow routine silence events to
dominate the replay distribution.

The replay procedure preserves every active-UE $\TX$ transition and
subsamples only redundant active-$\NT$ transitions when required by the
per-epoch replay budget. Furthermore, whenever both action classes are
present in replay memory, DQN minibatches are sampled approximately
equally from $\TX$ and $\NT$ transitions. The same replay rule is used
for both the sidelink-assisted and local-only configurations.

A nonterminal experience from epoch $t$ is not inserted until the true
decision state of epoch $t+1$ has been generated. Hence,
$\mathbf{s}(t+1)$ contains the realized next-epoch arrivals, fresh
channel observation, updated one-epoch-delayed peer information, and a
new contention-token realization. The replay tuple therefore preserves
the Bellman transition
\vspace{-4pt}\begin{equation} \vspace*{-4pt}
\left(
\mathbf{s}(t),
a(t),
r(t),
\mathbf{s}(t+1)
\right)
\end{equation}
rather than constructing an artificial next state from information
belonging to epoch $t$.

% =========================================================
% EXPLORATION
% =========================================================

\subsection{Load-Aware Exploration}
\label{subsec:exploration}

Applying conventional independent $\varepsilon$-greedy exploration to
every UE causes the amount of random access to increase with population
size. In a large active population, this can create excessive
exploration-induced collisions even when the learned policy itself is
well coordinated.

To avoid this behavior, $\varepsilon(t)$ is interpreted as a
epoch-level exploration intensity. Let
\vspace{-8pt}\begin{equation} \vspace*{-4pt}
N_{\mathrm{active}}(t)
=
|\mathcal{A}_B(t)|.
\end{equation}
Each active UE independently explores with probability
\begin{equation} \vspace*{-4pt}
p_{\mathrm{exp}}(t)
=
\min
\left\{
1,\,
\frac{
2\varepsilon(t)
}{
N_{\mathrm{active}}(t)
}
\right\}.
\label{eq:load_aware_exploration}
\end{equation}

An exploring UE selects uniformly from $\{\NT,\TX\}$. When the
clipping in \eqref{eq:load_aware_exploration} is inactive,
\begin{equation} \vspace*{-4pt}
\begin{aligned}
\mathbb{E}
\left[
N_{\mathrm{T}}^{\mathrm{exp}}(t)
\right]
&=
N_{\mathrm{active}}(t)
p_{\mathrm{exp}}(t)
\Pr
\{
\TX\mid\mathrm{explore}
\}
\\
&=
N_{\mathrm{active}}(t)
\frac{
2\varepsilon(t)
}{
N_{\mathrm{active}}(t)
}
\frac{1}{2}
\\
&=
\varepsilon(t).
\end{aligned}
\label{eq:expected_exploratory_tx}
\end{equation}

Thus, the expected number of random exploratory transmissions remains
approximately independent of the UE population rather than growing
linearly with the number of active terminals.

% =========================================================
% REWARD
% =========================================================

\subsection{Cooperative Anti-Capture Reward Design}
\label{subsec:reward-function}

Avoiding collisions alone is insufficient, as repeatedly serving the same UEs can leave others backlogged, reducing goodput and fairness. The reward therefore combines individual and cooperative components:

\vspace{-12pt}\begin{equation} \vspace*{-4pt}
r_i(t)
=
r_i^{\mathrm{ind}}(t)
+
\mathbbm{1}
\left\{
i\in\mathcal{A}_B(t)
\right\}
\omega_{\mathrm{team}}
R_{\mathrm{team}}(t),
\label{eq:final_reward}
\end{equation}
where $r_i^{\mathrm{ind}}(t)$ is the UE-specific reward and
$R_{\mathrm{team}}(t)$ is a common access-outcome term applied to all
backlogged UEs. The same reward design is used for both configurations.

\subsubsection{Successful-Service Utility}

Suppose UE $i^\star$ is the unique transmitter and is successfully
decoded. Its normalized delivered-goodput utility is
\begin{equation} \vspace*{-4pt}
g_{i^\star}(t)
=
\log
\left(
1+
\frac{
D_{i^\star}(t)
}{
D_{\mathrm{tar}}
}
\right),
\qquad
D_{\mathrm{tar}}
=
\lambda\Delta t,
\label{eq:goodput_utility}
\end{equation}
where normalization by $D_{\mathrm{tar}}$ measures service relative to
one epoch of offered traffic, while the logarithm limits sensitivity to
large service events.

The normalized pre-service backlog and waiting age are
\begin{equation} \vspace*{-4pt}
q_{i^\star}(t)
=
\operatorname{clip}
\left(
\frac{
\widetilde B_{i^\star}(t)
}{
B_{\max}
},
0,1
\right),
\label{eq:queue_reward_feature}
\end{equation}
and
\begin{equation} \vspace*{-4pt}
w_{i^\star}(t)
=
1-
\exp
\left(
-\frac{
W_{i^\star}(t)
}{
\tau_W
}
\right).
\label{eq:wait_reward_feature}
\end{equation}

To discourage service capture, the successful UE is weighted by its
relative backlog and waiting-age urgency among backlogged UEs:
\vspace{-8pt}\begin{equation} \vspace*{-4pt}
\rho_B(i^\star,t)
=
\frac{
\displaystyle
\sum_{j\in\mathcal{A}_B(t)}
\mathbbm{1}
\left\{
\widetilde B_j(t)
\leq
\widetilde B_{i^\star}(t)
\right\}
}{
|\mathcal{A}_B(t)|
},
\label{eq:backlog_percentile}
\end{equation}
and
\begin{equation} \vspace*{-4pt}
\rho_W(i^\star,t)
=
\frac{
\displaystyle
\sum_{j\in\mathcal{A}_B(t)}
\mathbbm{1}
\left\{
W_j(t)
\leq
W_{i^\star}(t)
\right\}
}{
|\mathcal{A}_B(t)|
}.
\label{eq:wait_percentile}
\end{equation}
The corresponding anti-capture urgency factor is
\begin{equation} \vspace*{-4pt}
u_{i^\star}(t)
=
\sqrt{
\rho_B(i^\star,t)
\rho_W(i^\star,t)
}.
\label{eq:urgency_weight}
\end{equation}
Thus, full success credit is favored when the transmitting UE is
relatively urgent in both backlog and waiting age, limiting repeated
service to recently served UEs with small fresh arrivals.

The resulting success utility is
\vspace{-6pt}\begin{equation} \vspace*{-4pt}
\begin{aligned}
R_{i^\star}^{\mathrm{succ}}(t)
=
u_{i^\star}(t)
\Big[
&
\omega_g
g_{i^\star}(t)
+
\omega_q
q_{i^\star}(t)
\\
&+
\omega_w
w_{i^\star}(t)
\Big],
\end{aligned}
\label{eq:success_reward}
\end{equation}
where $\omega_g$, $\omega_q$, and $\omega_w$ weight delivered goodput,
queue pressure, and waiting age, respectively.

% =========================================================
% INDIVIDUAL REWARD
% =========================================================

\subsubsection{UE-Specific Access Reward}

Let $\mathcal{S}(t)$ denote a successfully decoded singleton
transmission, and let $\mathcal{T}(t)$ denote the set of UEs selecting
the transmit action at decision epoch $t$. The UE-specific reward is
\begin{equation} \vspace*{-4pt}
r_i^{\mathrm{ind}}(t)
=
\begin{cases}
R_{i^\star}^{\mathrm{succ}}(t),
&
i=i^\star,\;
\mathcal{S}(t),
\\[1mm]

-R_{\mathrm{col}},
&
i\in\mathcal{T}(t),\;
N_{\mathrm{T}}(t)>1,
\\[1mm]

-R_{\mathrm{fail}},
&
i=i^\star,\;
N_{\mathrm{T}}(t)=1,\;
\neg\mathcal{S}(t),
\\[1mm]

-R_{\mathrm{idle}},
&
i\in\mathcal{A}_B(t),\;
N_{\mathrm{T}}(t)=0,
\\[1mm]

-R_{\mathrm{starve}},
&
\substack{
i\neq i^\star,\;
i\in\mathcal{A}_B(t),\\
\mathcal{S}(t),\;
W_i(t)>W_{i^\star}(t),
}
\\[2mm]

0,
&
\text{otherwise}.
\end{cases}
\label{eq:individual_reward}
\end{equation}

Thus, collisions, unsuccessful singleton attempts, and idle epochs with
backlog are penalized. After a successful singleton transmission, a
backlogged silent UE is penalized only if it has waited longer than the
successful UE; otherwise, silence is neutral.

\subsubsection{Cooperative Access-Outcome Reward}

The individual reward is complemented by a common team term:
\begin{equation} \vspace*{-4pt}
R_{\mathrm{team}}(t)
=
\begin{cases}
R_{\mathrm{team}}^{\mathrm{succ}},
&
\mathcal{S}(t),
\\[1mm]

-R_{\mathrm{team}}^{\mathrm{col}},
&
N_{\mathrm{T}}(t)>1,
\\[1mm]

-R_{\mathrm{team}}^{\mathrm{idle}},
&
N_{\mathrm{T}}(t)=0
\text{ and }
|\mathcal{A}_B(t)|>0,
\\[1mm]

-R_{\mathrm{team}}^{\mathrm{fail}},
&
N_{\mathrm{T}}(t)=1
\text{ and }
\neg\mathcal{S}(t).
\end{cases}
\label{eq:team_reward}
\end{equation}

The same $R_{\mathrm{team}}(t)$ is assigned to all backlogged UEs during
centralized training, encouraging singleton access while preserving
queue and waiting-aware credit assignment. This global access-outcome
signal is used only for reward construction during training and is not
required during decentralized inference; each UE still independently
selects $\mathrm{NT}$ or $\mathrm{T}$ from its local observation.

% =========================================================
% INFERENCE
% =========================================================

\subsection{Greedy Decentralized Inference}
\label{subsec:inference}

After training, the DQN parameters are frozen and no replay insertion,
Bellman update, or gradient calculation is performed. No team-reward or global access-outcome signal is required at inference. UE~$i$ selects
\begin{equation} \vspace*{-4pt}
a_i(t)
=
\arg\max_{a\in\mathcal{A}_i}
Q_{\boldsymbol{\theta}}
\left(
\mathbf{s}_i(t),a
\right),
\label{eq:greedy_inference}
\end{equation}
while an empty-buffer UE is forced to choose $\NT$.

Although the DQN decision is greedy for a fixed observation, the
contention token $\xi_i(t)$ continues to be randomly regenerated.
Consequently, the decentralized system retains stochastic
symmetry-breaking during greedy inference.

During inference, the local-only configuration uses only the local
observation $\mathbf{x}_i(t)$, whereas the sidelink-assisted
configuration augments the local information with the delayed
peer-state representation and coordination features defined in
\eqref{eq:sl_state}. No deterministic top-rank access rule is applied;
each UE independently evaluates its learned action-value function using
its available observation.

\vspace*{-5pt}

\section{Experiments}
\label{sec:testbed_design}

% =========================================================
% EXPERIMENT SETTINGS
% =========================================================

\subsection{Experiment Settings}
\label{subsec:experiment_setting}

The proposed MARL framework is evaluated using a population-driven LEO
D2S uplink simulator incorporating orbital dynamics, location-dependent
propagation, pedestrian UE mobility, finite-buffer traffic, and
decentralized shared-channel access, consistent with
Section~\ref{sec:system_model}.

Satellite positions and velocities are obtained from CelesTrak TLEs and
propagated using SGP4, with the catalog filtered to the considered
500--600~km LEO shell. A $72\times22$ Walker-Delta constellation at
550~km is retained as a fallback when TLE information is unavailable.
The receive beam of the satellite is directed toward the service-region center.

Two environments are considered: Suburban Erding, Germany, and Rural
Mistelbach District, Austria. Their similar representative latitudes
limit latitude-induced visibility differences while providing distinct
population densities and propagation conditions.

For the main evaluation, a $5\times5$~km$^2$ service region is used.
The active UE population follows \eqref{eq:population_ue_count}, assuming
one candidate handheld per resident and a $1.5\%$ activity factor,
yielding $61$ Suburban and $22$ Rural UEs. Thus, the contention load is
determined by the deployment environment rather than selected
independently. For the population-scalability study in 
Section~\ref{subsec:scalability_results}, the service area and
propagation environment are fixed while the number of simultaneously
active UEs is varied to isolate the effect of increasing contention load.

A separate train-and-test run is performed for each environment while
retaining the same MARL architecture and access model. Both observation
configurations have fixed input dimensions independent of $U$: five
features for Vanilla-IRIS and 34 features for A2A-IRIS.

Table~\ref{tab:location_scenarios} summarizes the location-dependent
parameters, while Table~\ref{tab:sim_params} reports the common orbital,
PHY, traffic, and learning settings.

\begin{table}[!t]
\caption{Location-Specific Scenario Parameters}
\vspace{-5pt}
\label{tab:location_scenarios}
\centering
\footnotesize
\renewcommand{\arraystretch}{1.10}
\setlength{\tabcolsep}{4pt}
\resizebox{\columnwidth}{!}{%
\begin{tabular}{l|cc}
\toprule
\rowcolor{gray!10}
\textbf{Parameter}
& \textbf{Suburban}
& \textbf{Rural} \\
\midrule
Reference location
& Erding, Germany
& Mistelbach District, Austria \\
Representative latitude
& $48.31^\circ$ N
& $48.57^\circ$ N \\
Representative longitude
& $11.91^\circ$ E
& $16.58^\circ$ E \\
Population density
& $162.7$ persons/km$^2$~\cite{erding_density}
& $59.6$ persons/km$^2$~\cite{mistelbach_density} \\
Service-area population
& $\approx4068$
& $\approx1490$ \\
Active UE density
& $2.44$ UEs/km$^2$
& $0.88$ UEs/km$^2$ \\
Active UEs, $U$
& $61$
& $22$ \\
Shadow-correlation time
& $20$ s
& $40$ s \\
Rain-rate parameter, $R_{0.01}$
& $80$ mm/h
& $50$ mm/h \\
Rain variability, $\sigma_{\ln R}$
& $1.26$
& $1.10$ \\
Rician $K$ offset
& $5$ dB
& $8$ dB \\
Rician $K$ elevation slope
& $0.20$
& $0.25$ \\
Minimum Rician $K$
& $3$ dB
& $5$ dB \\
\bottomrule
\end{tabular}%
}
\vspace{-10pt}
\end{table}

\begin{table}[!t]
\centering
\caption{Key Simulation and Learning Parameters.}
\label{tab:sim_params}
\renewcommand{\arraystretch}{1.06}
\setlength{\tabcolsep}{2.2pt}
\scriptsize

\begin{tabularx}{\columnwidth}{
@{}
>{\raggedright\arraybackslash}X
>{\raggedright\arraybackslash}p{0.18\columnwidth}
>{\raggedright\arraybackslash}X
>{\raggedright\arraybackslash}p{0.18\columnwidth}
@{}
}

\toprule

% ============================================================
% GROUP 1 + GROUP 2
% ============================================================

\multicolumn{2}{c}{\textit{Orbital and Scenario}}
&
\multicolumn{2}{c}{\textit{UE and Satellite RF}}
\\

\cmidrule(lr){1-2}
\cmidrule(lr){3-4}

\textbf{Parameter} & \textbf{Value}
&
\textbf{Parameter} & \textbf{Value}
\\
\midrule

Primary geometry
& TLE + SGP4
&
Duplex mode
& FDD UL
\\

TLE group / altitude
& STARLINK / 500--600 km
&
Carrier frequency, $f_c$
& 2 GHz
\\

Max. propagated satellites
& 1584
&
UE transmit power
& 23 dBm
\\

Fallback geometry
& Walker, $72\times22$, 550 km, $53^\circ$
&
UE antenna gain
& 0 dBi
\\

Service area
& $5\times5$ km$^2$
&
UE system loss
& 0 dB
\\

Candidate mapping, $\eta_{\mathrm{cand}}$
& 1 UE/person
&
UE EIRP
& 23 dBm
\\

Activity factor, $\eta_{\mathrm{act}}$
& 1.5\%
&
Satellite Rx gain
& 30 dBi
\\

UE mobility
& 3 km/h
&
Aperture diameter
& 2 m
\\

Min. elevation, $\theta_{\min}$
& $5^\circ$
&
Satellite $G/T$
& 1.1 dB/K
\\

\midrule

% ============================================================
% GROUP 3 + GROUP 4
% ============================================================

\multicolumn{2}{c}{\textit{Doppler and PHY}}
&
\multicolumn{2}{c}{\textit{Traffic and Access}}
\\

\cmidrule(lr){1-2}
\cmidrule(lr){3-4}

\textbf{Parameter} & \textbf{Value}
&
\textbf{Parameter} & \textbf{Value}
\\
\midrule

Doppler pre-compensation
& Enabled
&
Decision epoch, $\Delta t$
& 0.5 s
\\

Residual ICI penalty
& Enabled
&
Offered UL traffic per UE, $\lambda$
& 100 kbps
\\

Velocity-error std.
& 0.01 m/s
&
Traffic per epoch
& 50 kbit/UE
\\

GNSS position error
& 3 m
&
Buffer capacity, $B_{\max}$
& 4 Mbit
\\

Frequency quantization
& 10 Hz
&
Contention domain
& Shared UL
\\

Doppler-error correlation time
& 60 s
&
Action space
& $\{\NT,\TX\}$
\\

Nominal carrier bandwidth
& 10 MHz
&
Collision condition
& $N_{\mathrm T}>1$
\\

Shared UL BW, $B_{\mathrm{UL}}$
& 8.64 MHz
&
&
\\

Subcarrier spacing, $\Delta f$
& 30 kHz
&
&
\\
UL PRBs
& 24
&
&
\\
Max. HARQ attempts
& 4
&
&
\\

Target residual BLER
& 0.10
&
&
\\

Min. MCS SINR
& $-6$ dB
&
&
\\

\midrule

% ============================================================
% GROUP 5
% ============================================================

\multicolumn{4}{c}{\textit{Training Parameters}}
\\
\cmidrule(lr){1-4}

\textbf{Parameter} & \textbf{Value}
&
\textbf{Parameter} & \textbf{Value}
\\
\midrule

Training episodes
& 10000
&
Epochs per episode
& 500
\\

Hidden layers
& 128, 128
&
Local input dim.
& 5
\\

SL-assisted input dim.
& 34
&
Learning rate
& $10^{-3}$
\\

Replay capacity
& $10^5$
&
Min. replay size
& 2000
\\

Minibatch size
& 256
&
Target update
& 500 steps
\vspace*{-100pt}
\\
\bottomrule
\end{tabularx}
\end{table}
% \vspace{-4pt}
% =========================================================
% PROPAGATION
% =========================================================

\subsection{Propagation Environments}
\label{subsec:propagation_environments}

The suburban and rural scenarios use the same channel model from
Section~\ref{subsec:channel_model}, differing only in environment-specific
parameters. The simulator follows the corresponding 3GPP TR~38.811
S-band Suburban/Rural configuration~\cite{3gpp_tr38811} for
elevation-dependent LOS probability, LOS/NLOS shadow fading, and NLOS
clutter loss, using the nearest reference elevation in
$\{10^\circ,20^\circ,\ldots,90^\circ\}$.

LOS links use elevation-dependent Rician fading with the corresponding
NTN-TDL multipath profile, while NLOS links use Rayleigh fading with the NLOS profile. Free-space and atmospheric losses, rain attenuation,
satellite receive-beam gain, thermal noise, and residual Doppler-induced
ICI are also included.

GNSS-aided Doppler pre-compensation is assumed; hence, the full
geometry-induced Doppler is not treated as residual carrier offset.
Instead, the remaining frequency error is modeled as a temporally
correlated process accounting for GNSS position uncertainty, satellite
velocity uncertainty, and frequency quantization.

The access model, action space, traffic process, decision interval,
MCS/BLER mapping, and collision rule are identical across both scenarios.
Thus, their performance differences arise from population-driven
contention and environment-dependent propagation rather than algorithmic
changes.

% =========================================================
% TRAINING AND EVALUATION
% =========================================================
\subsection{Training and Evaluation Configuration}
\label{subsec:training_configuration}

The proposed MARL framework is instantiated in two configurations:
\emph{Vanilla-IRIS} and \emph{A2A-IRIS}. Both use parameter-shared
DQNs with the same architecture, action space, optimizer, reward,
exploration, replay mechanism, and training budget. Vanilla-IRIS uses the features in Section~\ref{subsec:state_action},
whereas A2A-IRIS additionally incorporates the
one-epoch-delayed peer-state histogram and current contention and
payload-priority ranks.

Load-aware $\varepsilon$-greedy exploration follows
Section~\ref{subsec:exploration}. Event-preserving replay follows
Section~\ref{subsec:dqn_training}, retaining all active-UE $\TX$
transitions while subsampling redundant $\NT$ transitions. Nonterminal
transitions are stored only after the realized next decision state is
available.

Steady-state performance is evaluated over 600 decision epochs using the
training traffic model. A separate inference-only buffer-drain test
initializes each UE with a 4-Mbit backlog, disables new arrivals, and
runs the frozen DQNs greedily ($\varepsilon=0$) for 15 epochs without
replay or parameter updates.

% =========================================================
% REFERENCES
% =========================================================

\subsection{Upper Bound and Reference Methods}
\label{subsec:baselines}
The two IRIS configurations, A2A-IRIS and Vanilla-IRIS, are evaluated
against an ideal upper bound (UB), three centralized channel-quality-based schedulers
(D0, D1, and D2), and two conventional random-access schemes.
A2A-IRIS and Vanilla-IRIS use the same parameter-shared MARL framework
and differ only in sidelink-assisted peer context, isolating the benefit
of peer-assisted coordination.

\noindent$\bullet$ \textbf{Upper-Bound Oracle (UB):}
The UB is a centralized current-CSI oracle that selects, in each epoch,
the UE with the largest queue-limited deliverable payload,
\begin{equation}\vspace*{-4pt}
C_i^{\mathrm{UB}}(t)=
\min\!\left\{
\widetilde{B}_i(t),
\frac{\eta_i^{\mathrm{SE}}\!\left(\gamma_i^{\mathrm{eff}}(t)\right)
B_{\mathrm{UL}}\Delta t\,\eta_{\mathrm d}}
{N_{r,i}(t)}
\right\}.
\end{equation}
Here, $\widetilde B_i(t)$ is the post-arrival UE backlog, $\eta_i^{\mathrm{SE}}(\gamma_i^{\mathrm{eff}}(t))$
is the SINR-dependent spectral efficiency, and $N_{r,i}(t)$ is the
required HARQ-round count; $B_{\mathrm{UL}}$, $\Delta t$, and
$\eta_{\mathrm d}$ denote the uplink bandwidth, epoch duration, and
delay-efficiency factor, respectively. The oracle assumes collision-free
access and no residual PHY failure, and its throughput is
\begin{equation}\vspace*{-4pt}
T_{\mathrm{UB}}
=
\frac{\sum_t \max_i C_i^{\mathrm{UB}}(t)}
{T\Delta t},
\end{equation}
where $T$ is the number of evaluation epochs.

\noindent$\bullet$ \textbf{Centralized Channel-Quality-Based Schedulers (D0, D1, and D2):}
These collision-free baselines use centralized single-UE scheduling and
differ only in channel-quality age. For delay $d\in\{0,1,2\}$, the estimated service
opportunity of backlogged UE~$i$ is
\begin{equation} \vspace*{-4pt}
\widehat C_i^{(d)}(t)
=
\frac{
\eta_i^{\mathrm{SE}}\!\left(\gamma_i^{\mathrm{eff}}(t-d)\right)
B_{\mathrm{UL}}\Delta t\,\eta_{\mathrm d}
}{
\widehat N_{r,i}\!\left(\gamma_i^{\mathrm{eff}}(t-d)\right)
},
\label{eq:delayed_service_metric}
\end{equation}
with
\begin{equation} \vspace*{-4pt}
i_d^\star(t)
=
\arg\max_{i:\widetilde B_i(t)>0}
\widehat C_i^{(d)}(t).
\label{eq:delayed_scheduler}
\end{equation}
D0 uses current effective SINR, while D1 and D2 use one- and two-epoch-old
effective SINR. Transmission is then evaluated using the current PHY
state, including MCS, HARQ, BLER, and payload. Thus, D0 is a current-SINR
centralized reference, whereas D1/D2 capture stale channel-quality loss.
D0 remains distinct from the UB because it retains stochastic decoding
and queue/HARQ effects.

\noindent$\bullet$ \textbf{$p^\star$-ALOHA:}
Each backlogged UE independently transmits with probability
\begin{equation} \vspace*{-4pt}
p^\star=\frac{1}{N_{\mathrm{active}}},
\end{equation}
where $N_{\mathrm{active}}$ is the number of non-empty UEs. This provides
a load-aware uncoordinated reference
\cite{10.1145/1478462.1478502,4548143}.

\noindent$\bullet$ \textbf{BEB:}
Binary exponential backoff (BEB)~\cite{840210} uses a contention window
(CW) that determines the range from which each UE draws a random backoff
counter, with $CW_{\min}=4$ and $CW_{\max}=256$. A UE transmits when its
counter expires; collisions double the CW up to $CW_{\max}$, while
success resets it to $CW_{\min}$.

\noindent$\bullet$ \textbf{Vanilla-IRIS (Ablation):}
Vanilla-IRIS uses the same two-layer DQN, binary action space, reward,
and training budget as A2A-IRIS, but observes only five local features:
normalized backlog, current effective SINR, waiting age, previous outcome, and the
UE-specific coordination token. It excludes the peer-state histogram and
relative coordination ranks available to A2A-IRIS.

All schemes use the same satellite geometry, traffic and finite-buffer
models, propagation assumptions, residual-Doppler treatment, MCS/BLER
mapping, and HARQ abstraction. D0--D2 use centralized collision-free
selection, while $p^\star$-ALOHA, BEB, Vanilla-IRIS, and A2A-IRIS share
the same contention-domain MAC model.
% =========================================================
% PERFORMANCE METRICS
% =========================================================

\subsection{Performance Metrics}
\label{subsec:performance_metrics}

The evaluation uses complementary metrics covering access efficiency,
reliability, fairness, resource cost, and computational overhead.

\noindent$\bullet$ \textbf{Effective Throughput/Goodput:}
The aggregate uplink goodput is
$\eta_{\mathrm{sys}}
=
\frac{1}{T\Delta t}
\sum_{t=1}^{T}
\sum_{i=1}^{U} D_i(t),$
where $D_i(t)$ is the successfully delivered payload, $T$ is the number
of evaluation epochs, and $\Delta t$ is the epoch duration.

\noindent$\bullet$ \textbf{Access Outcome Rates:}
The success rate is the fraction of epochs with a successfully decoded
singleton transmission, while the collision rate is the fraction with
multiple simultaneous transmissions. Idle and singleton PHY-failure
rates are also reported.

\noindent$\bullet$ \textbf{Fairness:}
Inter-user fairness is measured using Jain's index
\cite{digital1984quantitative},
% \begin{equation}
$J
=
\frac{
\left(\sum_{i=1}^{U}\eta_i\right)^2
}{
U\sum_{i=1}^{U}\eta_i^2
},$
% \label{eq:jain}
% \end{equation}
where $\eta_i$ is the delivered throughput of UE~$i$. Per-UE throughput
and served-UE statistics are additionally used to assess access capture.

\noindent$\bullet$ \textbf{Energy and Transmission-Resource Cost:}
RF efficiency is measured in joules per delivered Mbit, together with
transmission/HARQ round-equivalents per delivered Mbit. These capture
the cost of failed attempts and retransmissions beyond aggregate goodput.

\noindent$\bullet$ \textbf{Computational Complexity:}
Policy overhead is characterized by trainable DQN parameters,
forward-pass MAC operations per UE decision, and measured inference time.
Because the DQN is shared across UEs, parameter count is independent of
$U$, while aggregate neural inference cost scales linearly with $U$.

Queue saturation, transmission attempts, and served-UE fraction are
reported as diagnostic statistics where needed for interpretation.

% =========================================================
% SIMULATION SCOPE
% =========================================================

\subsection{Practicality of Real-World Evaluation}
\label{subsec:practicality}

Real-world evaluation of decentralized LEO D2S uplink access requires authorized access to an operational satellite link, compatible UEs, and coordination with the associated satellite and ground infrastructure. Such access is generally unavailable for controlled many-UE experiments,
particularly when the underlying radio-access decisions must be modified or observed.

Accordingly, IRIS is evaluated using a system-level simulation platform that combines TLE/SGP4 orbital dynamics, population-driven traffic, 3GPP TR~38.811-aligned S-band propagation~\cite{3gpp_tr38811},
pedestrian UE mobility, GNSS-aided Doppler pre-compensation, residual ICI, MCS-dependent stochastic BLER, finite-buffer dynamics, and bounded
HARQ. This enables controlled and reproducible evaluation under
time-varying LEO geometry, contention, queueing, and PHY conditions.

Operational LEO systems such as SpaceX Starlink and Amazon Leo rely
on operator-controlled radio-access infrastructure, limiting access for
external experimental modification. The simulation therefore captures
the D2S propagation, mobility, queueing, contention, and scheduling
characteristics relevant to the proposed access problem within the
system-level abstraction of Section~\ref{sec:system_model}.

{\color{black}
\section{Experiment Results}
\label{sec:results}

We evaluate IRIS against the baselines in
Section~\ref{subsec:baselines} using the setup of
Section~\ref{subsec:experiment_setting}. After training, the DQN
policies are frozen and evaluated greedily ($\epsilon=0$), with no
further adaptation.
Suburban and Rural scenarios are evaluated separately to preserve the
effects of their distinct UE densities and propagation conditions.
% The primary comparison is between A2A-IRIS and the local-only Vanilla-IRIS ablation. 
A2A-IRIS adds compact peer context and current contention and
payload-priority information, while all other learning, traffic, MAC,
reward, and PHY settings remain identical, thereby isolating the
benefit of sidelink-assisted A2A coordination.

%==============================================================
\subsection{Throughput Comparison}
\label{subsec:throughput_results}
As shown in Fig.~\ref{fig:effective_throughput}, A2A-IRIS achieves
2.1579~Mbps and 5.4478~Mbps in the Rural and Suburban scenarios,
corresponding to 99.9\% and 96.1\% of the respective UBs.

In comparison, D1 and D2 achieve 65.8--66.0\% of the UB in Rural and
41.7--42.4\% in Suburban. A2A-IRIS improves throughput over D1 and D2
by 51.7\% and 51.3\% in Rural, and 126.7\% and 130.4\% in Suburban,
respectively.

Without peer-state information, Vanilla-IRIS achieves only 1.5326~Mbps
in Rural and 1.2818~Mbps in Suburban, corresponding to A2A-IRIS gains
of 40.8\% and 325.0\%. A2A-IRIS also substantially outperforms
$p^\star$-ALOHA and BEB, demonstrating the benefit of compact peer
context for decentralized UE-side access.
\begin{figure}[t!]
    \centering
    \includegraphics[width=\columnwidth]{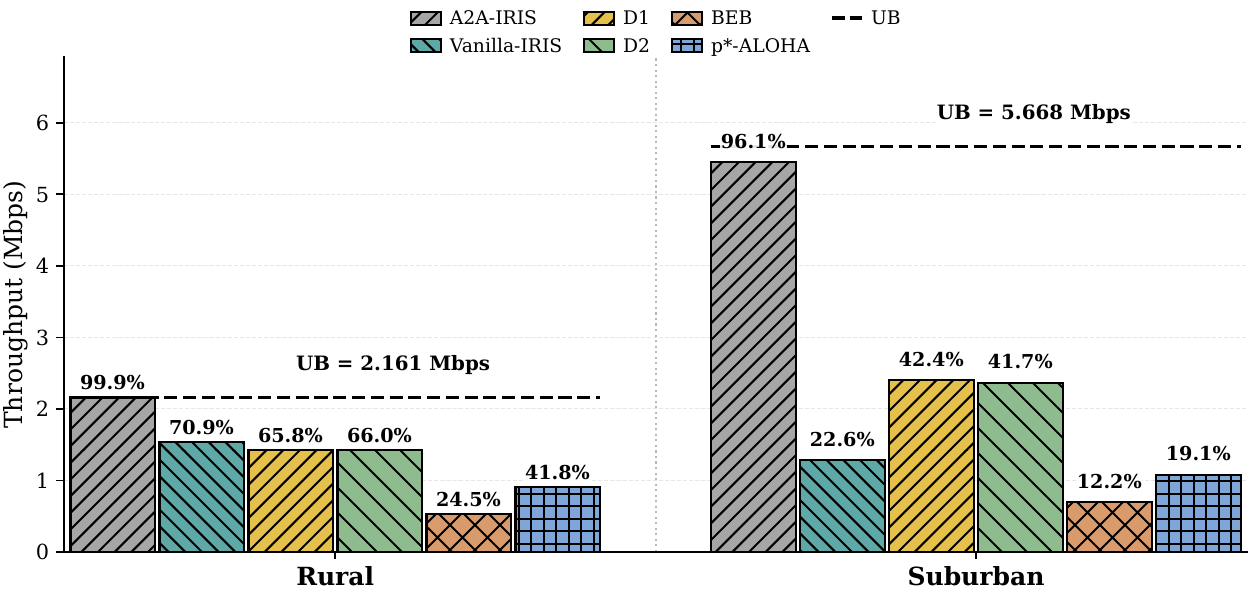}
    \vspace*{-18pt}
    \caption{Effective throughput of the considered schemes in the Rural and Suburban scenarios. The dashed lines denote the scenario-specific upper bounds, while the percentages above the bars indicate the fraction of the corresponding UB achieved by each scheme.}
    \label{fig:effective_throughput}
    \vspace*{-15pt}
\end{figure}
%%%%%%%%%%%%%%%%%%%%%%%%%%%%%%%%%%%%%%%%%%%%%%%%%%%%%%%%%%%
\subsection{Access Outcomes and Coordination Behavior}
\label{subsec:access_results}
Figs.~\ref{fig:success_rate} and~\ref{fig:collision_rate} further
illustrate the access behavior of the decentralized schemes. A2A-IRIS
achieves successful-epoch rates of 0.9244 and 0.8686 in the Rural and
Suburban scenarios, respectively, while maintaining collision rates of
0.0000 and 0.0433. In comparison, Vanilla-IRIS exhibits substantially
higher collision rates of 0.2735 and 0.1678, demonstrating the role of
sidelink information in coordinating distributed transmission decisions.
\begin{figure}[t!]
    \centering
    \includegraphics[width=\columnwidth]{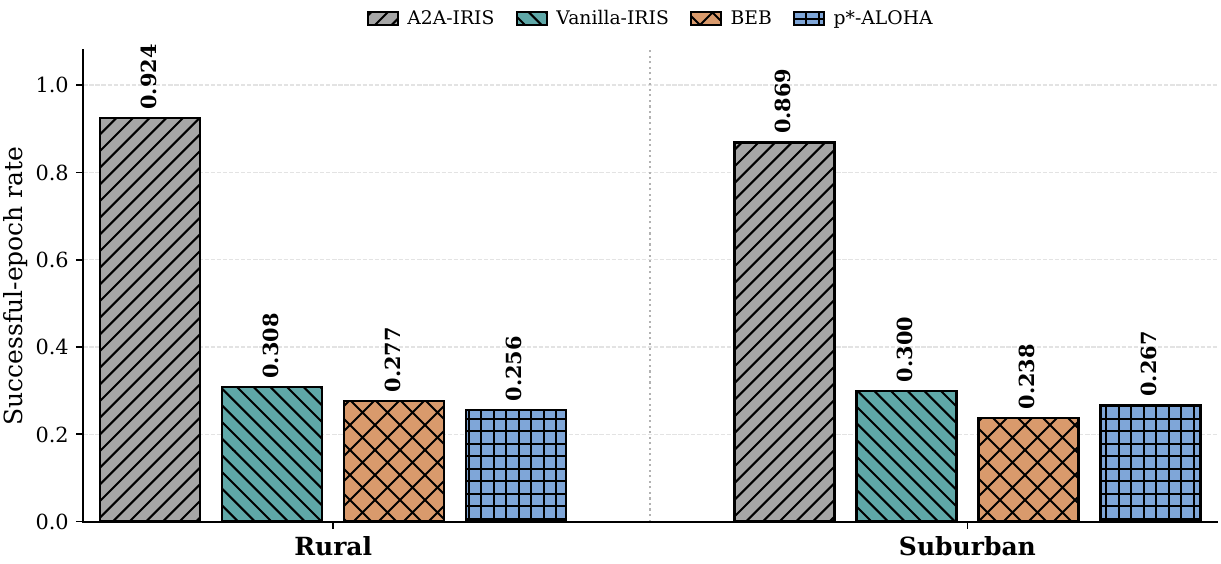}
    \vspace*{-18pt}     \caption{Successful-epoch rate of the decentralized access schemes under the Rural and Suburban scenarios.}
    \label{fig:success_rate}
        \vspace*{-10pt}
\end{figure}
\begin{figure}[t!]
    \centering
    \includegraphics[width=\columnwidth]{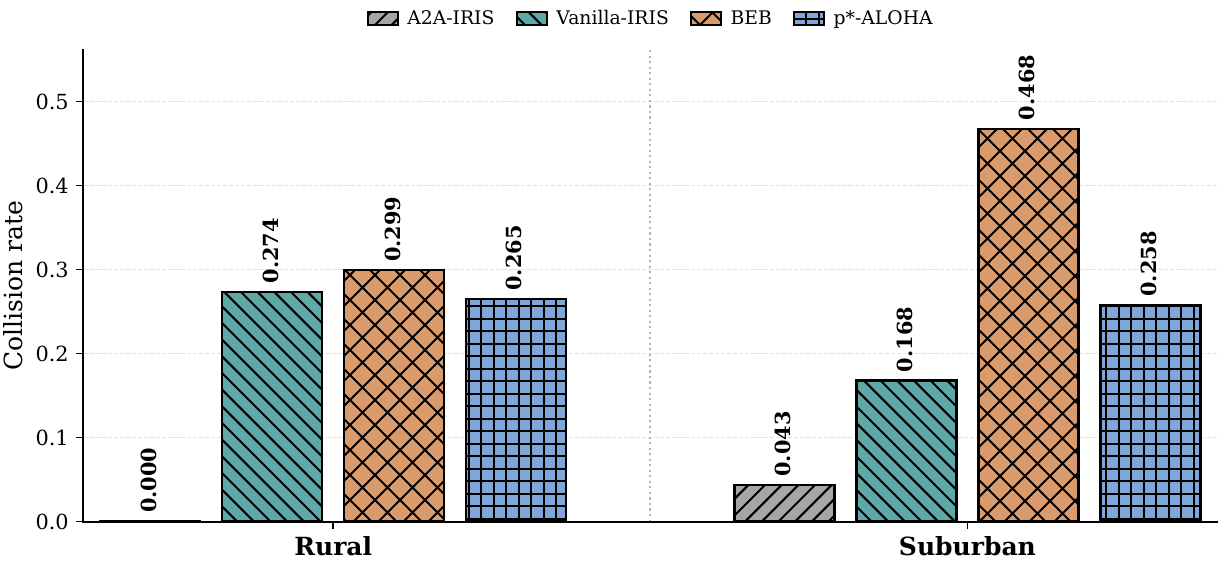}
    \vspace*{-18pt}     \caption{Collision rate of the decentralized access schemes under the Rural and Suburban scenarios.}
        \vspace*{-10pt}
    \label{fig:collision_rate}
\end{figure}

%---------------------------------------------------------------------------------
\subsection{Fairness, Queueing, and Persistent UE Capture}
\label{subsec:fairness_results}

Fig.~\ref{fig:fairness} compares the seed-wise Jain fairness of the
evaluated schemes. A2A-IRIS achieves the highest fairness in both
environments, reaching 0.9999 in Rural and 0.9954 in Suburban.
Vanilla-IRIS obtains 0.9789 and 0.8338, while D1 and D2 achieve
approximately 0.90--0.91 in Rural and 0.79 in Suburban. These results
show that A2A-IRIS provides more balanced service across UEs, with a
larger fairness advantage in the denser Suburban scenario.

The complementary queueing and service-coverage results in
Fig.~\ref{fig:queue} further illustrate this behavior. In Rural,
A2A-IRIS reduces queue saturation from 0.236 to 0.000, while both
schemes achieve a served-UE fraction of 1.000. In Suburban, the
served-UE fraction increases from 0.982 to 1.000, while queue
saturation decreases from 0.687 to 0.052. Overall, A2A-IRIS maintains
broader UE coverage in Suburban while substantially reducing queue
buildup in both environments.
\begin{figure}[t!]
    \centering
    \includegraphics[width=0.8\columnwidth]{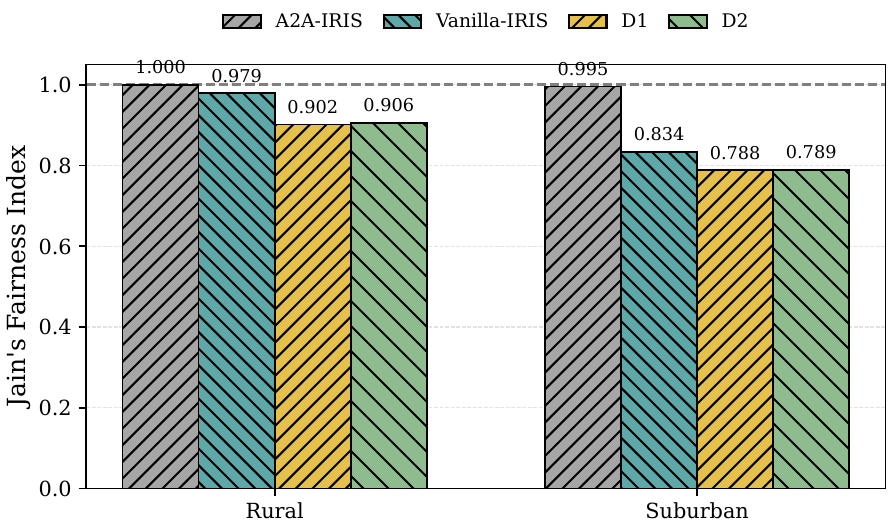}
    \vspace*{-5pt}     \caption{Jain's fairness index of A2A-IRIS, Vanilla-IRIS, and the delayed
    centralized references under the Rural and Suburban scenarios. The
    reported values are obtained by computing Jain's index for each
    evaluation seed and then averaging across seeds.}
        \vspace*{-15pt}
    \label{fig:fairness}
\end{figure}

\begin{figure}[t!]
    \centering
    \includegraphics[width=\columnwidth]{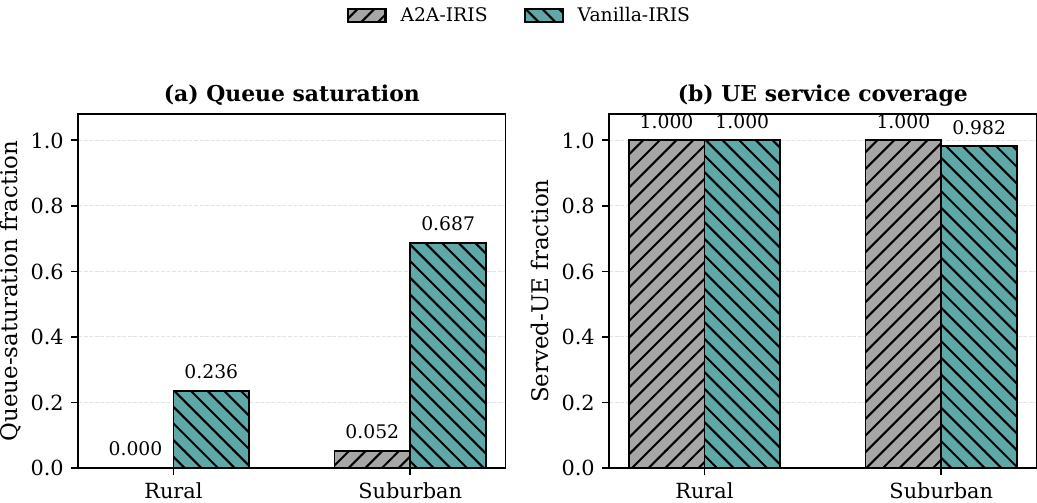}
    \vspace*{-15pt}     \caption{Queueing performance and UE service coverage of A2A-IRIS and Vanilla-IRIS in Rural and Suburban scenarios. Panel~(a) shows the queue-saturation fraction, while panel~(b) shows the served-UE fraction; lower saturation indicates reduced backlog accumulation, whereas higher coverage indicates broader UE service.}
        \vspace*{-15pt}
    \label{fig:queue}
\end{figure}

\subsection{Sensitivity to Delayed CSI}
\label{subsec:delay_results}
Fig.~\ref{fig:delay_sensitivity} shows the impact of CSI aging on the
centralized scheduler. The $D=0$ case is the instantaneous-CSI reference,
while $D=1$ and $D=2$ represent stale-CSI operation. In the Rural scenario,
throughput decreases from 1.7293~Mbps at $D=0$ to 1.4222~Mbps and
1.4266~Mbps at $D=1$ and $D=2$, respectively. Similarly, in the Suburban
scenario, it drops from 3.2979~Mbps to 2.4028~Mbps and 2.3643~Mbps.
In contrast, A2A-IRIS achieves 2.1579~Mbps and 5.4478~Mbps in the Rural
and Suburban scenarios, respectively, approaching the corresponding UB
while outperforming the centralized delayed-CSI cases.
\begin{figure}[t!]
    \centering
    \includegraphics[width=\columnwidth]{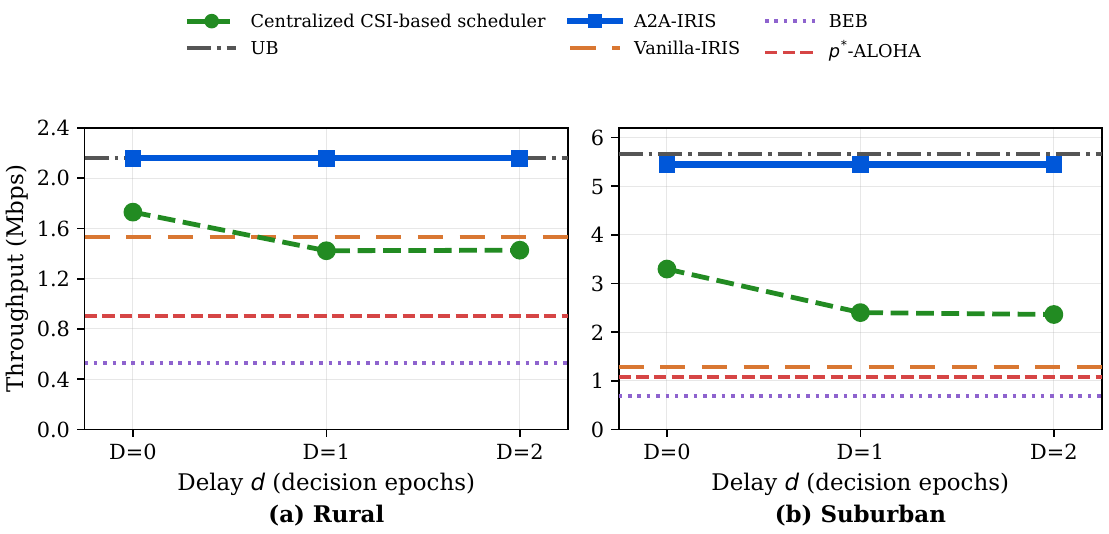}
    \vspace*{-15pt}     \caption{Impact of CSI delay on the centralized scheduler in the Rural and
Suburban scenarios. The $D=0$ case denotes the instantaneous-CSI centralized reference, while $D=1$ and $D=2$ represent stale-CSI operation. Horizontal
lines denote the UB and the decentralized access schemes.}
    \vspace*{-12pt}
    \label{fig:delay_sensitivity}
\end{figure}

%%%%%%%%%%%%%%%%%%%%%%%%%%%%%%%%%%%%%%%%%%%%%%%%%%%%%%%%%%%%%%%%%%%%%%%%%%%%%%%%
\subsection{Energy and Transmission-Resource Efficiency}
\label{subsec:energy_results}

The efficiency results in Figs.~\ref{fig:energy_cost}
and~\ref{fig:resource_cost} show that the throughput improvement of
A2A-IRIS is not obtained through excessive transmission activity.
A2A-IRIS requires only 0.091 and 0.037~J/Mbit in the Rural and Suburban
scenarios, respectively, with corresponding transmission-resource costs
of 0.913 and 0.372 transmission/HARQ round-equivalents per Mbit. In contrast, the
substantially higher energy cost of Vanilla-IRIS is primarily caused by
uncoordinated simultaneous transmissions. Collided transmissions consume
RF energy without delivering payload, while the undelivered traffic
remains queued and must be attempted again in subsequent epochs.
Consequently, the combination of wasted collision energy and repeated
transmission attempts increases both the energy and transmission-resource
cost per successfully delivered Mbit. These results indicate that
sidelink-assisted coordination simultaneously improves effective
throughput and reduces unsuccessful transmission overhead.
\begin{figure}[t!]
    \centering
    \includegraphics[width=\columnwidth]{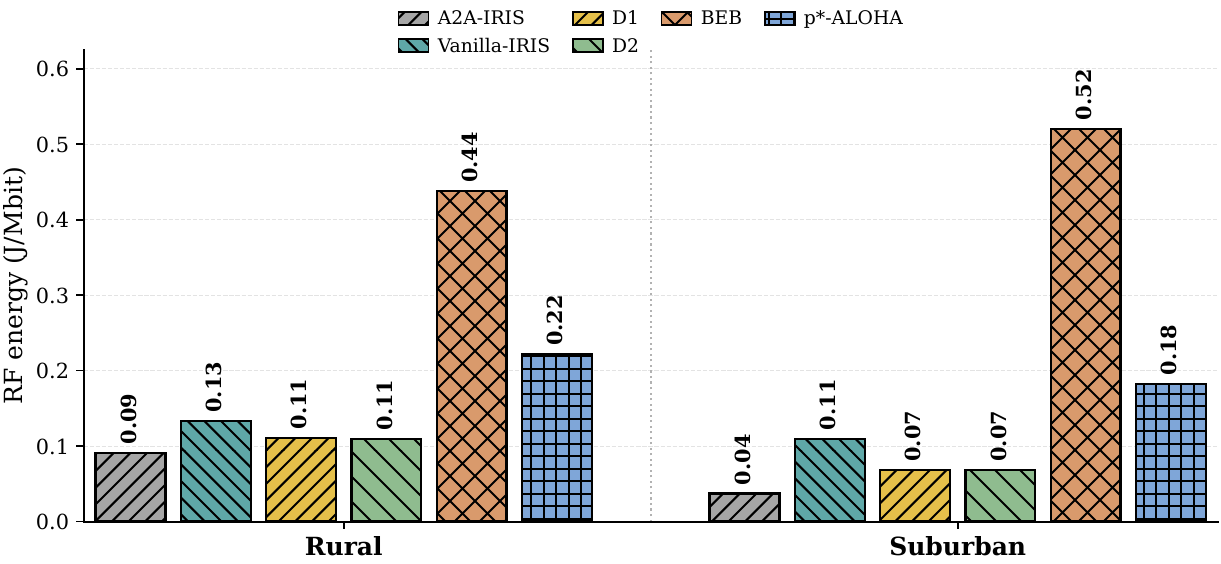}
    \vspace*{-15pt}     \caption{RF transmission energy cost per successfully delivered Mbit
under the Rural and Suburban scenarios. The higher cost of Vanilla-IRIS
results from collision-induced wasted transmissions and subsequent
reattempts of undelivered traffic.}
    \vspace*{-15pt}
    \label{fig:energy_cost}
\end{figure}
\begin{figure}[t!]
    \centering
    \includegraphics[width=\columnwidth]{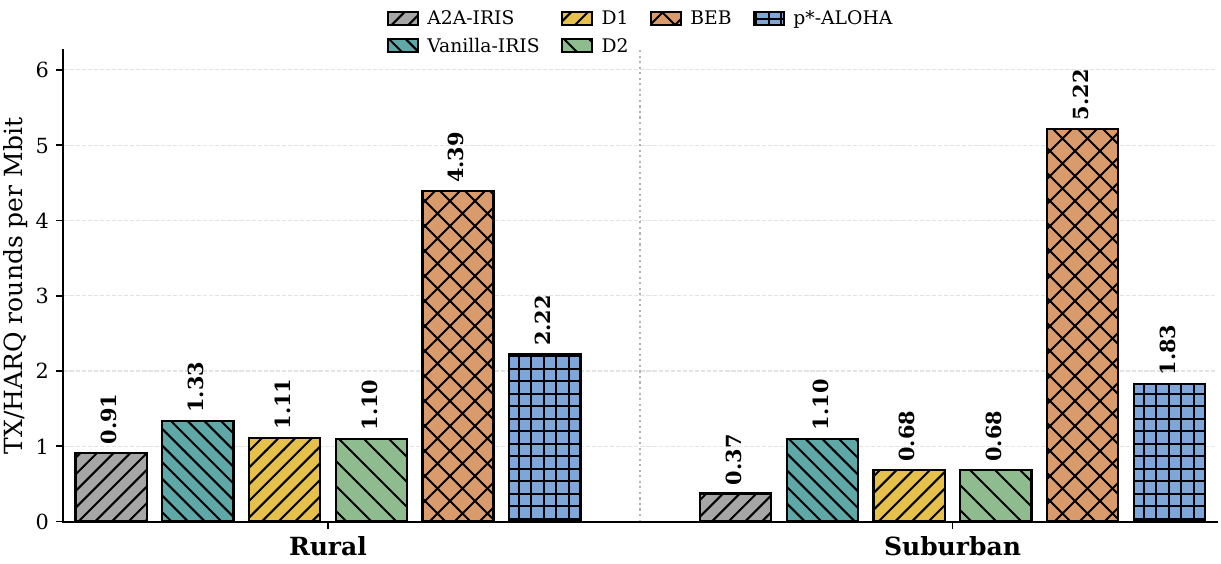}
    \vspace*{-15pt}     \caption{Normalized transmission-resource cost, measured in transmission/HARQ
round-equivalents per successfully delivered Mbit, under the Rural and
Suburban scenarios. The larger cost of Vanilla-IRIS reflects the additional
transmission attempts caused primarily by uncoordinated collisions.}
    \vspace*{-10pt}
    \label{fig:resource_cost}
\end{figure}

%%%%%%%%%%%%%%%%%%%%%%%%%%%%%%%%%%%%%%%%%%%%%%%%%%%%%%%%%%%%%%%%%%%%%%%%%%%%%
\subsection{Computational Complexity}
\label{subsec:complexity_results}

Fig.~\ref{fig:dqn_complexity} compares the inference complexity of
A2A-IRIS and Vanilla-IRIS. Using the same two-layer DQN backbone,
A2A-IRIS processes a richer sidelink-assisted observation and therefore
requires $21{,}250$ trainable parameters and $20{,}992$ MACs per UE,
compared with $17{,}538$ parameters and $17{,}280$ MACs for
Vanilla-IRIS, corresponding to increases of approximately $21.2\%$ and
$21.5\%$, respectively.

Despite this overhead, inference latency remains on the order of tens of
microseconds per UE. A2A-IRIS additionally requires only lightweight
histogram aggregation and rank computation, without iterative
optimization or matrix inversion. Because the DQN input dimension is
fixed, the per-UE forward-pass complexity remains $O(1)$ with respect to
the active UE population $U$, while the aggregate neural inference cost
scales as $O(U)$.
\begin{figure}[t!]
    \centering
    \includegraphics[width=\columnwidth]{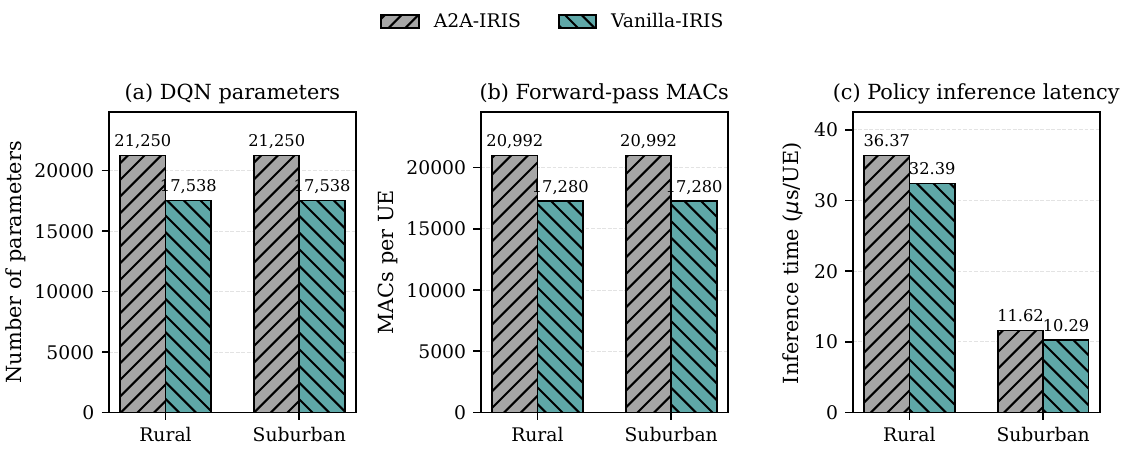}
    \vspace*{-15pt}     \caption{Computational complexity of A2A-IRIS and Vanilla-IRIS under the
Rural and Suburban scenarios. The comparison includes the DQN parameter
count, forward-pass MAC operations per UE, and measured policy inference
latency per UE.}
    \vspace*{-15pt}
    \label{fig:dqn_complexity}
\end{figure}
%%%%%%%%%%%%%%%%%%%%%%%%%%%%%%%%%%%%%%%%%%%%%%%%%%%%%%%%%%%%%%%%%%%%%%%%%%%

\subsection{Sensitivity to Active UE Load}
\label{subsec:scalability_results}

Fig.~\ref{fig:scalability} evaluates IRIS under increasing active-UE
load while keeping the service area and propagation environment fixed.
The active population is varied as
$U\in\{10,20,30,40,80,100\}$, while the nominal operating points remain
$U=22$ for Rural and $U=61$ for Suburban. Thus, the largest tested load
corresponds to approximately $4.5\times$ and $1.6\times$ the nominal
Rural and Suburban populations, respectively.

Across the evaluated range, A2A-IRIS achieves higher throughput than
Vanilla-IRIS and D1, with throughput increasing as the contention
population grows. Vanilla-IRIS and D1 exhibit weaker scaling due to the lack of peer awareness and stale channel information, respectively.
These results highlight the potential of A2A-IRIS for population-dense
deployments, where increasing active-UE populations create stronger
contention over the shared uplink and make peer-assisted coordination
increasingly beneficial. The sweep characterizes this behavior over the
evaluated range rather than extrapolating beyond the tested populations.

\begin{figure}[t!]
    \centering
    \includegraphics[trim={0 5mm 0 0},clip, width=\columnwidth]{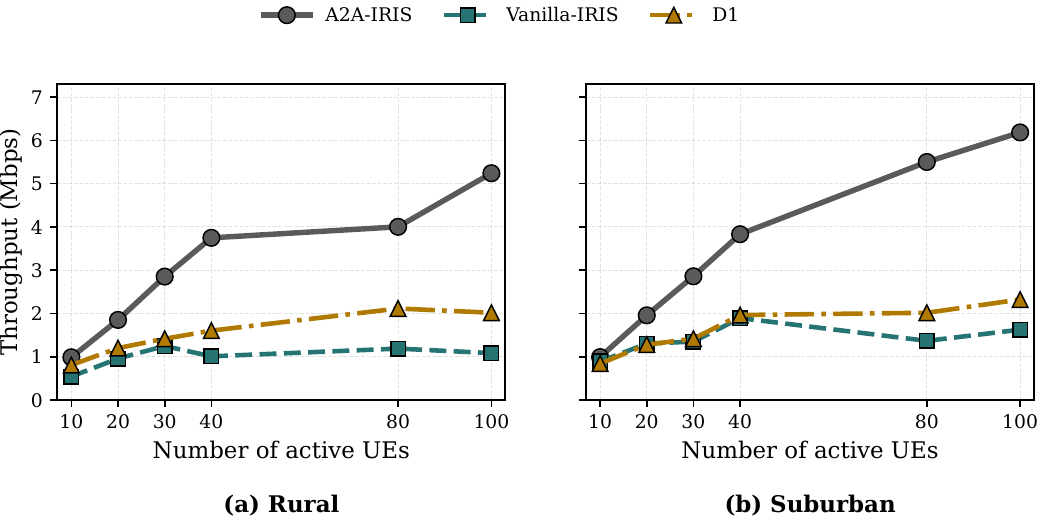}
    \vspace*{-18pt}     
    \caption{Fig. 12. Active-UE load scalability of A2A-IRIS, Vanilla-IRIS, and D1 with fixed service area and propagation environment. The active UE population is evaluated at 10, 20, 30, 40, 80, and 100 UEs, while the nominal main-evaluation points remain 22 UEs for Rural and 61 UEs for Suburban.}
    \label{fig:scalability}
        \vspace*{-18pt}  
\end{figure}

%%%%%%%%%%%%%%%%%%%%%%%%%%%%%%%%%%%%%%%%%%%%%%%%%%%%%%%%%%%%%%%%%
\subsection{Comparison with Existing NTN Random-Access}
\label{subsec:sota_comparison}
Table~\ref{tab:sota_comparison} compares IRIS with representative NTN
random-access approaches. Existing schemes mainly address contention,
multi-user reception, activity detection, or cooperative reception,
typically without explicit peer-state information at the transmitting
UEs. In contrast, A2A-IRIS performs UE-side access decisions using
compact sidelink-assisted peer context to reduce uncoordinated
transmissions. The Vanilla-IRIS ablation isolates this contribution by
retaining the same MARL framework without peer-assisted coordination.
Thus, A2A-IRIS is distinguished by integrating compact peer context with
grant-free decentralized access control.
\begin{table*}[t!]
\centering
\renewcommand{\arraystretch}{1.15}
\caption{Architectural Comparison of IRIS with Representative NTN
Random-Access Approaches.}
\vspace*{-5pt}
\label{tab:sota_comparison}
\resizebox{\textwidth}{!}{%
\begin{tabular}{lcccccc}
\hline\hline
\textbf{Approach}
& \textbf{Grant-Free}
& \textbf{UE-Side Decision}
& \textbf{Learning}
& \textbf{Peer Context}
& \textbf{Concurrent Access Handling}
& \textbf{Primary Focus} \\
\hline
Lee \textit{et al.}~\cite{9844988}
& Yes
& Yes
& MARL
& No
& Collision based
& Decentralized RA \\

Yu \textit{et al.}~\cite{10132304}
& Yes
& Yes
& DRL
& No
& NOMA/SIC
& LEO IoT RA \\

Shen \textit{et al.}~\cite{9849120}
& Yes
& No
& No
& No
& Multi-user detection
& Activity/channel estimation \\

Ying \textit{et al.}~\cite{10.1109/JSAC.2023.3273699}
& Yes
& No
& No
& Multi-satellite
& Cooperative reception
& Quasi-synchronous RA \\

Vanilla-IRIS (Ablation)
& Yes
& Yes
& MARL (shared DQN)
& Local only
& Collision based
& Learned contention \\

\textbf{A2A-IRIS (Proposed)}
& \textbf{Yes}
& \textbf{Yes}
& \textbf{MARL (shared DQN)}
& \textbf{Compact sidelink}
& \textbf{Collision based}
& \textbf{Peer-assisted contention}\\
\hline\hline
\end{tabular}
 \vspace*{-20pt}}
\end{table*}

%%%%%%%%%%%%%%%%%%%%%%%%%%%%%%%%%%%%%%%%%%%%%%%%%%%%%%%%%%%%%%%%%%%%%%%%%%%
\subsection{Satellite Mobility and Handover Implications}
\label{subsec:handover_discussion}

LEO orbital motion creates time-varying geometry that affects slant range,
elevation, path loss, antenna gain, and Doppler. We capture these
variations using CelesTrak TLE trajectories propagated with SGP4 when
available, with a Walker constellation as fallback. For each
UE, the serving satellite is selected among visible satellites using the
highest reference received-power metric and may change as geometry
evolves.

Because the shared DQN is not satellite-specific, serving-satellite
changes are handled through the evolving SINR-related observations
without policy reinitialization. The study therefore models dynamic link
selection and PHY variation rather than full handover signaling.
Handover overhead and beam-sweep effects are outside the scope.

% \subsection{Overall Discussion}
% \label{subsec:results_discussion}
%==============================================================

% }
\vspace*{-5pt}
\vspace{1pt}
\section{Conclusion}
\label{sec:conclusions}

This work presented IRIS, a parameter-shared MARL framework for
decentralized grant-free LEO D2S uplink access, with A2A-IRIS as the
sidelink-assisted realization and Vanilla-IRIS as its local-only
ablation. A2A-IRIS achieves 2.1579 and 5.4478~Mbps in Rural and
Suburban scenarios, corresponding to 99.9\% and 96.1\% of the
scenario-specific UBs and gains of 40.8\% and 325.0\% over
Vanilla-IRIS. It also outperforms one- and two-epoch delayed centralized
schedulers by approximately 51--130\%, while improving fairness, service
coverage, and transmission efficiency with moderate computational
overhead. Active-UE load experiments further show sustained throughput
gains over the evaluated range. Overall, sidelink-assisted peer
coordination improves decentralized LEO D2S contention resolution and
resource efficiency. Future work will consider imperfect sidelink
delivery and overhead, dynamic handover and multi-satellite access,
richer action spaces, and implementation-oriented validation.

\bibliographystyle{IEEEtran} 
% \newpage
\bibliography{reference}

\end{document}